\documentclass[a4paper,fleqn,usenatbib]{mnras}
\usepackage{deluxetable}
\usepackage{footmisc}
\usepackage{graphicx}
\usepackage{amssymb}
\usepackage{amsmath}
\usepackage{wasysym}
\usepackage{color}
\usepackage[normalem]{ulem}
\usepackage{psfrag}
\usepackage{multirow,bigdelim}
\graphicspath{{./}}

\newcommand{\ohte}{\hbox{(O/H)$_{\rm T_e}$}}

\newcommand{\halpha}{\hbox{H$\alpha$}}
\newcommand{\hbeta}{\hbox{H$\beta$}}

\newcommand{\rtwothree}{\hbox{$R_{23}$}}

\newcommand{\ntwoion}{\hbox{[\ion{N}{ii}]~$\lambda$6583}}
\newcommand{\oaur}{\hbox{[\ion{O}{iii}]~$\lambda$4363}}
\newcommand{\fiveoo}{\hbox{[\ion{O}{iii}]~$\lambda$5007}}

\newcommand{\oaurratiotthree}{\hbox{[\ion{O}{iii}]~$\lambda$4363/$\lambda\lambda(4959+5007)$}}
\newcommand{\oaurratiottwo}{\hbox{[\ion{O}{ii}]~$\lambda$7320+7330$/\lambda\lambda(3726+3729)$}}
\newcommand{\sulratio}{\hbox{[\ion{S}{ii}]~$\lambda$6717/$\lambda$6731}}

\newcommand{\ttwothree}{\hbox{$T_2-T_3$}}

\newcommand{\mdotyr}{\hbox{$M_\odot$ yr$^{-1}$}}

\newcommand{\mstar}{\hbox{$M_{\star}$}}
\newcommand{\dssfr}{\hbox{$\Delta\log$(SSFR)}}
\newcommand{\teff}{\hbox{$T_{\rm eff}$}}
\newcommand{\mturn}{\hbox{$M_O$}}
\newcommand{\zmax}{\hbox{$Z_O$}}

\newcommand{\mzsfrrelation}{\hbox{$M_{\star}$--$Z$--${\rm SFR}$}}

\newcommand{\mpa}{\hbox{MPA/JHU}}   

\newcounter{minirefcount}
\newcommand{\msol}{M_{\odot}}

\title[Recalibrating Oxygen Abundance Diagnostics]{A Recalibration of Strong Line Oxygen Abundance Diagnostics via the Direct Method and Implications for the High Redshift Universe}

\author[J. S. Brown et al.]{
Jonathan S. Brown,$^{1}$\thanks{E-mail: brown@astronomy.ohio-state.edu}
Paul Martini,$^{1,2}$
and Brett H. Andrews$^{3}$
\\
$^{1}$Department of Astronomy, The Ohio State University, 140 West 18th Avenue, Columbus, OH 43210, USA\\
$^{2}$Center for Cosmology and Astro-Particle Physics, The Ohio State University, 191 West Woodruff Avenue, Columbus, OH 43210, USA\\
$^{3}$PITT PACC, Department of Physics and Astronomy, University of Pittsburgh, 3941 O'Hara Street, Pittsburgh, PA 15260, USA
}

\date{Accepted XXX. Received YYY; in original form ZZZ}

\pubyear{2015}

\begin{document}
\label{firstpage}
\pagerange{\pageref{firstpage}--\pageref{lastpage}}
\maketitle

\begin{abstract}
We use direct method oxygen abundances in combination with strong optical emission lines, stellar masses ($\mstar$), and star formation rates (SFRs) to recalibrate the N2, O3N2, and N2O2 oxygen abundance diagnostics. We stack spectra of $\sim$200,000 star-forming galaxies from the Sloan Digital Sky Survey in bins of $\mstar$ and SFR offset from the star forming main sequence ($\dssfr$) to measure the weak emission lines needed to apply the direct method. All three new calibrations are reliable to within $\pm 0.10$ dex from $\log(\mstar/\msol) \sim 7.5 - 10.5$ and up to at least $200~\msol$ yr$^{-1}$ in SFR. The N2O2 diagnostic is the least subject to systematic biases. We apply the diagnostics to galaxies in the local universe and investigate the $M_{\star}$--$Z$--${\rm SFR}$ relation. The N2 and O3N2 diagnostics suggest the SFR dependence of the $M_{\star}$--$Z$--${\rm SFR}$ relation varies with both $\mstar$ and $\dssfr$, whereas the N2O2 diagnostic suggests a nearly constant dependence on SFR. We apply our calibrations to a sample of high redshift galaxies from the literature, and find them to be metal poor relative to local galaxies with similar $\mstar$ and SFR. The calibrations do reproduce direct method abundances of the local analogs. We conclude that the $M_{\star}$--$Z$--${\rm SFR}$ relation evolves with redshift.   
\end{abstract}


\section{Introduction}
\label{sec:intro}
Galaxies are continually undergoing chemical enrichment. Gas is condensed into stars, processed into heavier elements, and returned to the interstellar medium. This gas, enriched by the products of stellar nucleosynthesis and/or supernova ejecta, is reincorporated into new generations of stars, where it is enriched once again. A galaxy may also accrete low metallicity gas from the intergalactic medium, which both dilutes the ISM and provides fuel for a new generation of stars to form. This interplay between star formation, chemical enrichment, and accretion of new material is a central component of galaxy evolution.   

An episode of star formation increases a galaxy's stellar mass and enriches the ISM. A substantial body of work has shown that there are correlations between stellar mass ($\mstar$), star formation rate (SFR), and gas phase oxygen abundance. The correlation between $\mstar$ and gas phase oxygen abundance is called the Mass-Metallicity Relation \citep[MZR;][]{Lequeux79, Tremonti04}. The MZR extends from low mass, extremely metal deficient galaxies like Leo P \citep{Skillman13} up to massive galaxies with 2-3 times the solar oxygen abundance \citep{Tremonti04, Moustakas11}. 

The MZR often serves as a benchmark for models of galaxy evolution because the details of the MZR are direct probes of the underlying physics. For instance, \citet{Tremonti04} describe how the shape of the MZR requires galactic winds to efficiently remove metals from low mass galaxies. Subsequent cosmological models \citep[e.g.][]{Dave06b,Oppenheimer06,Finlator08,Dave11a,Dave11b} incorporated winds into their cosmological models in order to better understand the origin of the MZR. In the context of their momentum driven wind models, the mass loading parameter $\eta \equiv \dot{M}_{\rm outflow}/\dot{M}_{\star}$ is proportional to the inverse of the velocity dispersion of the halo, which scales with the halo mass to the one third power, $\eta \propto 1/\sigma_h \propto M_h^{-1/3}$ \citep{Murray05, Oppenheimer06}. Once the star formation has reached an equilibrium with the inflowing and outflowing gas, the metallicity is $Z = y/(1+\eta)$ where $y$ is the effective yield. In the limit that $\eta \gg 1$, the slope of the MZR is ultimately related to how $\mstar$ scales with $M_h$, since $\log(Z) \propto -\log(\eta) \propto \frac{1}{3} \log(M_h)$.

There is good observational evidence for a second parameter that affects the relationship between $\mstar$ and $Z$ such that galaxies with higher star formation rates have lower metallicities at fixed stellar mass \citep[the $\mzsfrrelation$ relation;][]{Ellison08Apj, Mannucci10, LaraLopez10}. This relation is also apparent in high signal-to-noise ratio stacked spectra of SDSS galaxies \citep{Andrews13}. However, it has intriguingly not been seen in the CALIFA sample of 150 nearby galaxies studied with integral field spectroscopy by \citet{Sanchez13}.

The exact form of the SFR dependence is less clear, but if the fuel for star formation is lower metallicity gas accreted from the IGM, this would produce an anticorrelation between gas phase metallicity and SFR. The form of the secondary dependence of the MZR on SFR offers insights into several open questions, such as how star formation is regulated, and how the processes that govern galactic inflows and outflows operate in detail \citet{Dave11a,Dave11b,Lilly13}. 

In addition to the local MZR and its dependence on SFR, the same correlations can be studied in high redshift galaxies in order to probe galaxy formation and evolution in the early universe \citep{Shapley05, Erb06, Maiolino08, Steidel14, Zahid14b, Sanders15}. Furthermore, the correlation between $\mstar$, Z, and SFR in the early universe, and how that relates to the correlations observed in the local universe, constrains how the population of galaxies has evolved over cosmic time \citep{Zahid14,Maier14,Izotov15}.

Accurate and precise metallicity measurements are vital to gain physical insights from both local correlations and evolution over cosmic time. The most reliable oxygen abundances are determined with the ``direct method'', or ``$T_e$ method'' \citep{Dinerstein90}. Under the right conditions, the electron temperature of ionized gas can be directly measured from the temperature sensitive intensity ratios of collisionally excited forbidden lines (e.g. [\ion{O}{iii}]~$\lambda$4363/[\ion{O}{iii}]~$\lambda$5007). As oxygen is one of the primary coolants in the ISM, the temperature is anticorrelated with abundance. The density of the gas can be measured from density sensitive lines (e.g. $\sulratio$). For a given temperature and density the emissivity of a given ionic species can be computed, which can then be used to determine relative abundances. 

The direct method is subject to some biases. Temperature fluctuations and gradients in \ion{H}{ii} regions produce a bias towards lower metallicities \citep{Peimbert67,Kobulnicky99} . This bias also applies to integrated (as well as stacked) spectra of galaxies. Hotter regions have brighter auroral lines, which can bias the direct method toward higher electron temperatures and correspondingly lower metallicities. Additionally, the assumption of a Maxwell-Boltzmann electron energy distribution has recently come into question \citep{Nicholls12,Dopita13}. If electron energies are instead well described by a $\kappa$-distribution, this may contribute to the well known temperature discrepancy problem \citep{Garcia-Rojas06, Garcia-Rojas07,Nicholls12,Blanc15}, although this is less of a concern for relative comparisons of direct method abundances. Even with the potential for these systematic effects, the direct method is widely regarded as the standard for nebular abundances.

In practice, dectecting the auroral lines (e.g. $\oaur$) requires a significant investment of observational resources for even the brightest, most metal poor galaxies and \ion{H}{ii} regions. At present, most spectroscopy comes from low to moderate SNR, and direct method abundances are typically not practical.

In order to estimate the metallicities of galaxies without the use of the auroral lines, so-called 'strong-line' diagnostics were developed based on the more easily measured nebular emission lines \citep{Pagel79, Alloin79}. There have been many efforts to calibrate these diagnostics via theoretical \citep[e.g.,][]{McGaugh91, Zaritsky94, Dopita00, Charlot01, Kewley02, Kobulnicky04, Tremonti04, Stasinska06} and empirical means \citep[e.g.,][]{Pilyugin03, Pettini04, Pilyugin05, Pilyugin10, Pilyugin12, Marino13, Bianco15}.

Perhaps the most common of these diagnostics is $\rtwothree~\equiv~(\text{[\ion{O}{ii}]}~\lambda 3727 + \text{[\ion{O}{iii}]}~\lambda\lambda 4959, 5007) / \hbeta$ \citep{Edmunds84,McCall85,Dopita86,Zaritsky94}. $\rtwothree$ encodes some information about the overall oxygen abundance, but the ratio is ultimately determined by the excitation of the [\ion{O}{ii}] and [\ion{O}{iii}] lines. This leads to the double valued nature of $\rtwothree$, which complicates its use as an abundance diagnostic.

Fortunately there are other nebular lines that encode information about the gas phase oxygen abundance, and nitrogen is the most accessible of these. Nitrogen has both primary origin, where the amount of nitrogen produced in stars and returned to the ISM is independent of metallicity, and secondary origin, where the amount of nitrogen produced is proportional to metallicity \citep{Alloin79, Vila-Costa93, Considere00}. In the high metallicity regime, nitrogen is secondary and the nitrogen abundance increases faster than the oxygen abundance. Furthermore, some strong line ratios are temperature sensitive since, for instance, the [\ion{O}{ii}]~$\lambda$ 3727 \AA\ line requires a significantly higher energy to excite than the [\ion{N}{ii}]~$\lambda$6583 \AA\ line \citep{Pilyugin10}. As a result, nitrogen based diagnostics can serve as indicators of the oxygen abundance.

Many strong-line calibrations are often inconsistent with one another. \citet{Kewley08} show the extent to which the various strong line calibrations disagree and provide a framework for mapping one strong line metallicity onto another. Many of the strong-line calibrations differ simply because they use different calibration samples, but the situation is more complicated than sample selection. Some calibrations utilize grids from photoionization simulations \citep{McGaugh91,Zaritsky94,Kewley02}, while others use unique samples of \ion{H}{ii} regions \citep[e.g.,][]{Marino13} which themselves are often heterogeneous compilations of samples from the literature \citep[e.g.,][]{Pettini04,Pilyugin10}.

Empirical abundance diagnostics have the benefit of being calibrated on direct method measurements, but due to selection effects the calibration samples are often biased toward low metallicity \ion{H}{ii} regions \citep{Jones15}. The application of these calibrations to integrated spectra of moderately star forming galaxies requires significant extrapolation from the \ion{H}{ii} regions that compose most calibration samples. Furthermore, most empirical calibrations will result in erroneous metallicities if, for instance, the ionization conditions of the galaxies in question differ significantly from the calibration sample \citep{Dopita00,Kewley02,Steidel14}.

Recently, several studies have shown that stacking the spectra of a sufficiently large number of galaxies can boost the S/N of the auroral lines to a detectable level \citep{Liang07, Andrews13}. We use the stacking technique presented in \citet{Andrews13} to obtain direct method oxygen abundances for galaxies spanning a wide range in $\mstar$ and SFR. Our stacking method mitigates the potential for bias by binning galaxies we expect to have similar metallicities based on the small intrinsic scatter of the MZR and $\mzsfrrelation$ relation. We then recalibrate the popular strong line abundance diagnostics with the direct method oxygen abundances, and apply the new calibrations to data taken from the literature.

We adopt the following notation for the principal diagnostic emission line ratios: 

\begin{flalign*}
&\text{N2} = \text{[\ion{N}{ii}]}~\lambda6583 / \halpha & \\
&\text{O3N2} = \text{[\ion{O}{iii}]}~\lambda5007 / \hbeta / \text{[\ion{N}{ii}]}~\lambda6583 / \halpha & \\
&\text{N2O2} = \text{[\ion{N}{ii}]}~\lambda6583 / \text{[\ion{O}{ii}]}~\lambda3727 & \\
&R_2= \text{[\ion{O}{ii}]}~\lambda3727 / \hbeta & \\
&R_3= \text{[\ion{O}{iii}]}~\lambda\lambda4959,5007 / \hbeta & \\
&R_{23} = R_2+R_3 & \\
&P = R_3/R_{23} & \\
\end{flalign*}

Section \ref{sec:data} describes our selection and stacking process. Section \ref{sec:analysis} describes our empirical calibrations of \ohte. In Section \ref{sec:results} we present our newly derived calibrations. In Section~\ref{sec:discussion} we apply our calibrations to various samples of galaxies and discuss the implications for the $\mzsfrrelation$ relation. Finally, we briefly summarize our results in Section \ref{sec:conc}.

\section{Data}
\label{sec:data}

\subsection{Sample Selection}
\label{sec:sample}
Our sample of galaxies is derived from the SDSS Data Release 7 (DR7; \citet{Abazajian09}). We begin with the $\mpa$ catalog of galaxies with stellar masses \citep{Kauffmann03b}, SFRs \citep{Brinchmann04, Salim07}, and oxygen abundances \citep[][hereafter T04]{Tremonti04}. We discard AGN dominated galaxies with the standard Baldwin-Philips-Terlevich (BPT) diagram \citep{Baldwin81} and the criterion for star forming galaxies from \citet{Kauffmann03a}:

\begin{multline}
\log([\text{\ion{O}{iii}}]~\lambda5007 / \hbeta) < \\
   0.61[\log([\text{\ion{N}{ii}}]~\lambda6583/\halpha) - 0.05]^{-1} + 1.3.
\label{eq:sf}
\end{multline} 

Our S/N requirements are the same as those presented in \citet{Andrews13}. We restrict our sample to galaxies with \hbeta, \halpha, and $\ntwoion$ detected at $>5\sigma$. For galaxies with $\fiveoo$ detected at $>3\sigma$, we apply the selection criteria shown in Equation~\ref{eq:sf}. In order to include galaxies with high metallicity (and inherently weak $\fiveoo$) we include galaxies with $\fiveoo$ detected at $<3 \sigma$ but $\log(\ntwoion/\halpha) < 0.4$.

We also take significant care to inspect low mass galaxies ($\log[M_*] < 8.6$) and remove galaxies with poor photometric deblending (flagged with \verb|DEBLEND_NOPEAK| or \verb|DEBLENDED_AT_EDGE|) or otherwise spurious stellar mass determinations. These selection cuts leave a total of 208,529 galaxies in our sample.

We emphasize that a limitation of this analysis is that the data were obtained with single fibers centered on resolved galaxies, and therefore not all of the light is included in the 3\arcsec\ diameter fiber aperture. For reference, 3\arcsec\ corresponds to 2.2 kpc at the median redshift ($z = 0.078$) of our sample. The missing fraction due to this aperture bias will depend on redshift for galaxies of similar sizes, and will depend on mass and star formation rate due to the flux-limited nature of the sample. This aperture bias is important because galaxies exhibit radial abundance gradients \citep[e.g.][]{Searle71,Kennicutt03,Bresolin09a,Bresolin09b,Berg13,Sanchez14} that will cause abundances measured in the central region of a galaxy to overestimate the total abundance. \citet{Tremonti04} investigated this aperture bias for SDSS observations and found metallicity variations of 0.05 to 0.11 dex with redshift for galaxies of similar absolute $z-$band magnitudes. \citet{Kewley05} studied aperture effects with the Nearby Field Galaxy Survey and recommended that fiber spectroscopy include at least $>20$\% of the galaxy light (typically $z > 0.04$ for SDSS observations) to minimize systematic and random errors, and this corresponds to most of our sample. Based on these studies, we estimate that aperture biases are comparable to the scatter in the inferred O/H for galaxies of similar stellar mass and star formation rate.

Another limitation of single-fiber observations is they simply present an incomplete picture of the properties of galaxies. One example is that while \citet{Sanchez13} found a very tight relationship between integrated stellar mass and metallicity with integral field data from CALIFA \citep{Sanchez12}, they did not find any dependence of metallicity on star formation rate at fixed stellar mass. Another example is the analysis by \citet{Belfiore15} of nebular data for 14 galaxies with P-MaNGA, the prototype instrument for the ongoing MaNGA survey \citep{Bundy15}. Those authors found a substantial spread in O/H values at fixed N/O for regions within individual galaxies, which is in contrast to the stronger correlation exhibited by the central regions from single-fiber observations.

\subsection{Stacking Procedure}
\label{sec:stack}

The auroral lines of [\ion{N}{ii}], [\ion{O}{ii}], and [\ion{O}{iii}] are generally weak and typically undetectable in most SDSS galaxy spectra. However, previous studies \citep[e.g.][]{Liang07, Andrews13} have demonstrated that stacking spectra to reduce the contribution of random fluctuations in the measured flux is a viable way to obtain sufficient S/N to measure the auroral lines.

The stacking method relies on the fact that the random noise in a composite spectrum of $N$ galaxies scales roughly as $1/\sqrt{N}$; it is advantageous for our bins to contain a large number of galaxies in order to reduce the noise in the spectrum as much as possible. However, we also want each bin to span a very small range in \textit{actual} (O/H) so that we are stacking qualitatively similar galaxies. The chosen bin widths are a compromise between these two goals. 

\begin{figure*}
\psfrag{O}[c][][1.]{$\odot$}
\psfrag{S}[c][][1.]{$\star$}
\centering{\includegraphics[scale=1.,width=\textwidth,trim=0.pt 0.pt 0.pt 0.pt,clip]{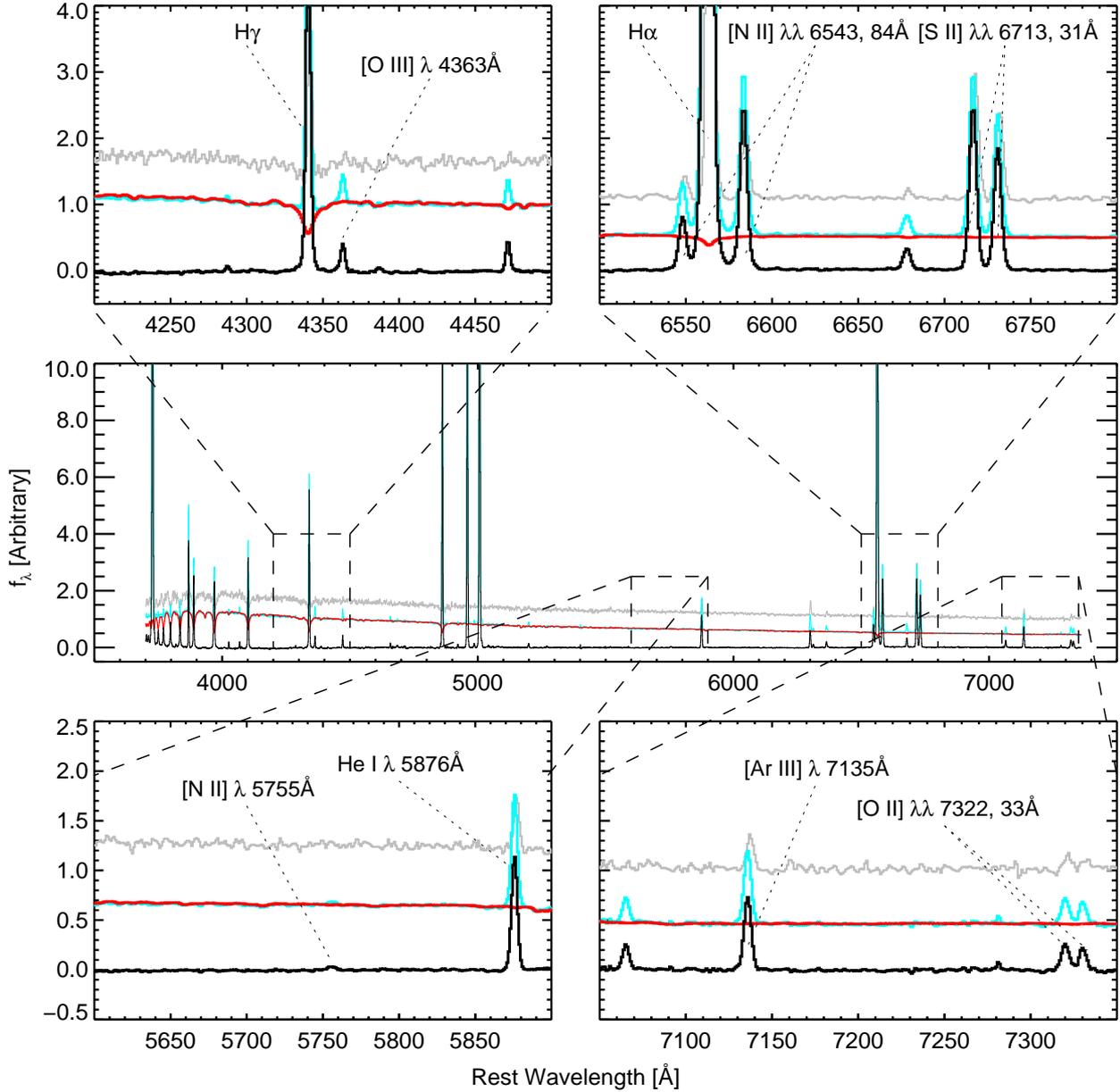}}
\caption{Illustration of how stacking improves the S/N of the weak lines. The top and bottom sets of plots show different regions of the middle spectrum. In each panel, the gray line shows the raw SDSS spectrum (shifted to rest frame wavelength), the blue line shows the stacked spectrum, the red line shows the fit to the stellar continuum, and the thick black line shows the spectrum after stellar continuum subtraction.}
\label{fig:ref_spec_OII}
\end{figure*}

Before stacking the spectra, we follow the same reduction process described in \citet{Andrews13}. Starting with the spectra that have been processed with the SDSS pipeline \citep{Stoughton02}, we correct for Galactic reddening using the extinction values from \citet{Schlegel98}. We then shift each spectrum to the rest frame using redshifts from the $\mpa$ catalog. We interpolate each spectrum onto a wavelength grid spanning 3700\AA--7360\AA\ with spacing $\Delta \lambda = 1$\AA. In order to compare galaxies at various distances we normalize each spectrum to the stellar continuum with the mean continuum flux from 4400\AA--4450\AA. Thus when we measure the line flux we effectively measure the equivalent width of the line. At fixed \mstar, normalizing to the stellar continuum is acceptable since the luminosities of the galaxies are essentially the same. Figure~\ref{fig:ref_spec_OII} demonstrates the benefit of stacking. In the raw SDSS spectrum of a single galaxy (gray line), the weak auroral lines are undetectable. They become fairly evident after stacking (blue line). After removing nearby stellar continuum features (red line), the previously undetectable auroral lines are prominent features in the final spectrum (black line).

\subsection{Choice of Stacking Parameters}
\label{sec:justify_stack}

Our goal is to derive improved strong line calibrations, so one of the parameters we use to assign galaxies to a stack is similar strong line ratios.  However, the strong line ratios show considerable dependence on more parameters than just metallicity, such as incident spectral shape, ionization parameter, and gas density \citep{Dopita00,Kewley02,Dopita13}. For example, Steidel et al. (2014) demonstrated that variations in line ratios due to a factor of five change in metallicity could be reproduced with only a factor of two change in ionization parameter. Figure 2 clearly demonstrates that there is a substantial range in stellar mass and star formation rate at a constant value of the N2, O3N2, N2O2, or R23 strong line diagnostics.

As in \citet{Andrews13}, we assume that galaxies with similar stellar masses and star formation rates have similar physical conditions, and therefore similar values of the other parameters that impact the connection between strong line ratio and metallicity. We consequently only stack galaxies with similar stellar masses and star formation rates to minimize the dispersion in galaxy properties in each stack. Good support for this approach comes from an investigation of stacking by \citet{Andrews13}. They compared electron temperatures and abundances for galaxies with individual auroral line detections to stacks of the same sample of galaxies and found good agreement within the measurement uncertainties.

We have performed a bootstrap analysis as an additional validation of this approach. For this analysis we chose four bins of different star formation rates at the same stellar mass. We resampled each bin 100 times and processed them with our analysis pipeline to derive the metallicity.  We found the median of the bootstrap metallicity distribution agreed well with the stack value for each bin. The spread in the metallicity distribution ($\sim 0.15$ dex) was somewhat larger than the formal metallicity uncertainties, but smaller than the variations in the strong line ratios at fixed stellar mass and star formation rate ($\sim 0.2$ dex).

We have chosen to use both stellar mass and star formation rate because there is good evidence that metallicity depends on star formation rate at fixed mass (Ellison et al. 2008; Mannucci et al. 2010; Lara-Lopez et al. 2010). In addition, we expect galaxies with different star formation rates at fixed mass may differ in other parameters (incident spectral shape, etc.).  While the integral field study by \citet{Sanchez13} did not find that metallicity depends on star formation rate at fixed mass, we emphasize that our decision to stack in both quantities is also motivated by how other physical parameters vary with star formation rate.

It is also well known that stellar mass and star formation rate are well correlated, a correlation known as the star forming main sequence \citep{Brinchmann04, Salim07, Noeske07, Whitaker12, Zahid12b, Kashino13}. In order to characterize this dependence, \citet{Salim14} showed that the parameter $\dssfr$

\begin{equation}
\dssfr = \log({\rm SSFR}) - \left\langle \log({\rm SSFR}) \right\rangle_{\small{M_{\star}}}
\label{eq:dssfr}
\end{equation}

\noindent
is more effective than both SFR and SSFR at identifying low and high oxygen abundance outliers across a wide range in $\mstar$. The quantity $\left\langle \log({\rm SSFR}) \right\rangle_{\small{M_{\star}}}$ is the median $\log$(SSFR) of galaxies at $\mstar$. Thus $\dssfr$ is defined relative to the star forming main sequence rather than an arbitrary value (e.g. 1 $\mdotyr$).

\begin{figure*}
\psfrag{O}[c][][1.]{$\odot$}
\psfrag{S}[c][][1.]{$\star$}
\centering{\includegraphics[scale=1.,width=0.45\textwidth,trim=0.pt 0.pt 190.pt 75.pt,clip]{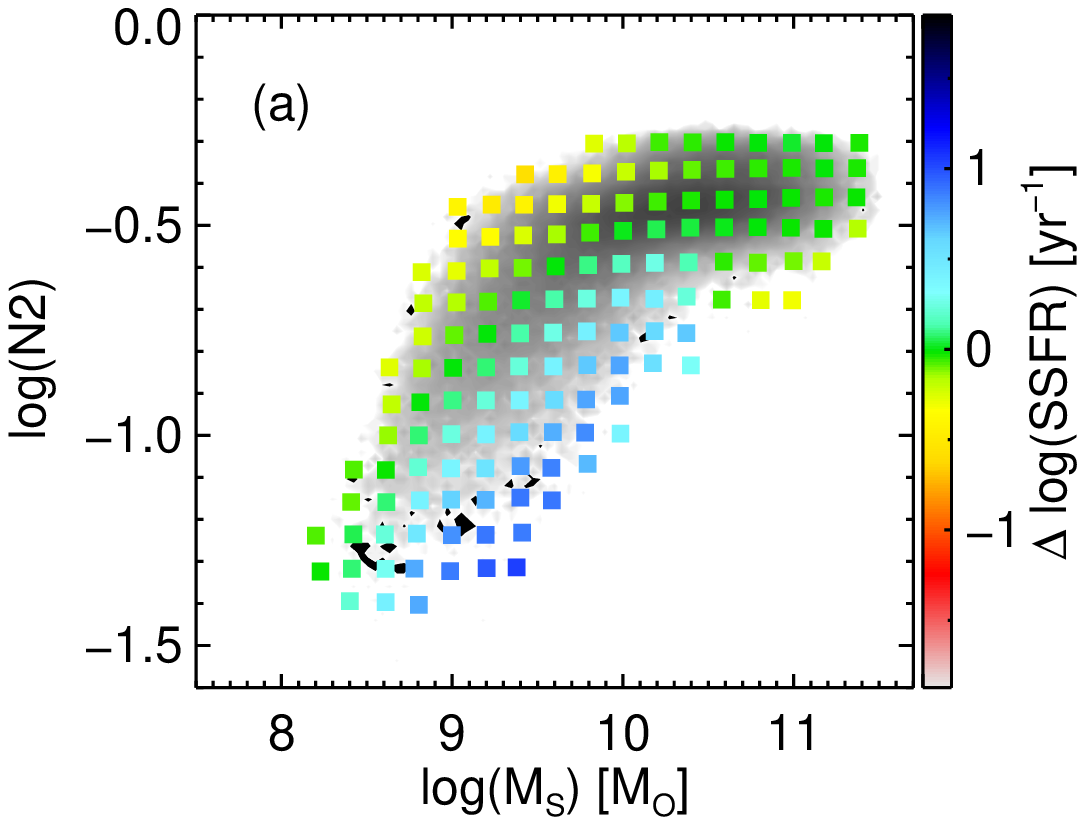}}
\centering{\includegraphics[scale=1.,width=0.45\textwidth,trim=0.pt 0.pt 190.pt 75.pt,clip]{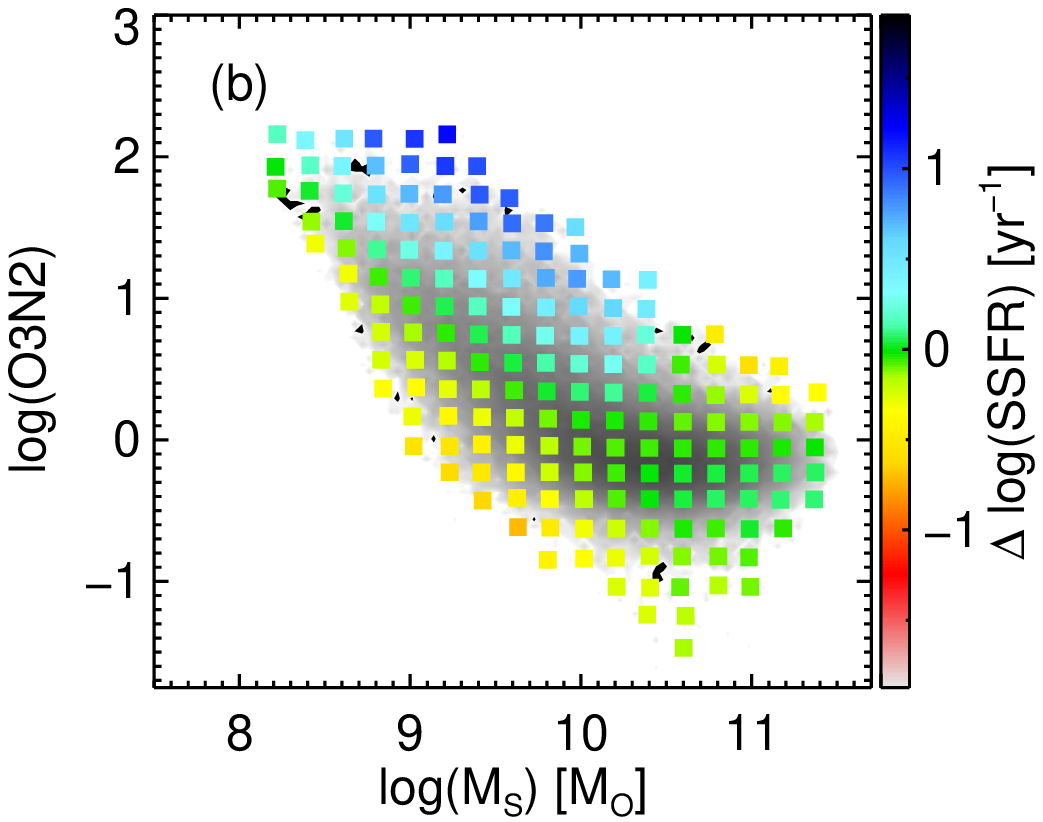}}

\centering{\includegraphics[scale=1.,width=0.45\textwidth,trim=0.pt 0.pt 190.pt 75.pt,clip]{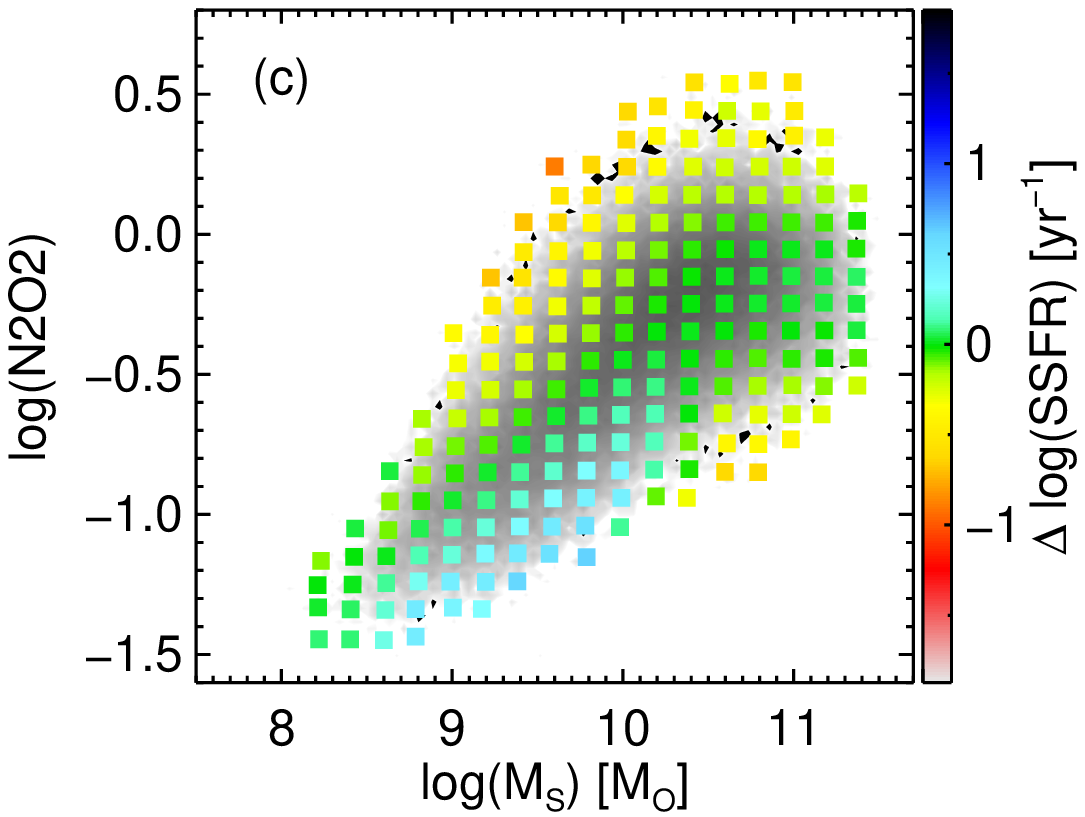}}
\centering{\includegraphics[scale=1.,width=0.45\textwidth,trim=0.pt 0.pt 190.pt 75.pt,clip]{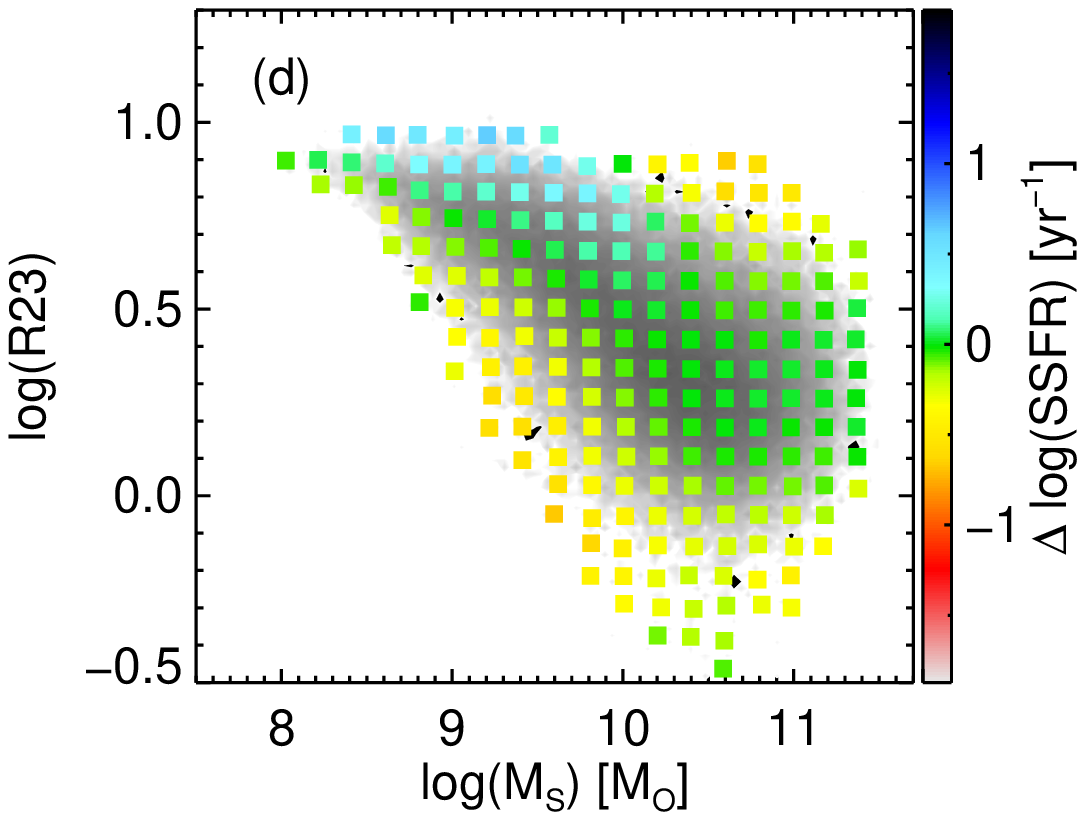}}
\caption{Distribution of SDSS galaxies in the various $\mstar$-diagnostic planes considered here. The individual galaxies are binned in a 2D grid. Color coding denotes the average $\dssfr$ of each bin; the underlying gray scale shows the relative density of galaxies in our input catalog. Top left to bottom right, the panels show N2, O3N2, N2O2, and $\rtwothree$ versus $\mstar$. In the case of N2, O3N2, and N2O2 the scatter in the diagnostic at fixed $\mstar$ and $\dssfr$ is generally small compared to the overall range spanned by the diagnostic. In the case of $\rtwothree$, the distribution of galaxies at fixed $\mstar$ and $\dssfr$ is rather broad compared to the range spanned by the diagnostic, making the $\rtwothree$ line ratios of a given stack less meaningful.} 
\label{fig:dssfr}
\end{figure*}

Binning in $\dssfr$ rather than SFR is also beneficial for calibrating the relationship between the strong line ratios and $\ohte$. Figure~\ref{fig:dssfr} shows that at a fixed strong line ratio, there is significant scatter in $\mstar$. Since $\mstar$ and SFR are correlated, absolute SFR does not necessarily correspond to a lower oxygen abundance at a fixed strong line value. Furthermore, since $\dssfr$ is a reflection of the SFR density, galaxies with similar $\dssfr$ ought to have similar ionization conditions. The same does not hold true for galaxies with similar SFR but different stellar masses, since a relatively low mass, compact star forming galaxy will have more intense ionization conditions than a more massive galaxy with relatively diffuse star formation.

Our choice of bin widths was largely \textit{ad hoc}. It is clear from Figure 11 of \citet{Andrews13} that there is some scatter in $\ohte$ at fixed $\mstar$ and $\dssfr$. Our primary motives were to (1) resolve the $\mzsfrrelation$ relationship, (2) include enough galaxies in metal rich stacks to measure $\ohte$, and (3) limit the total number of stacks to keep the stacking procedure, stellar continuum subtraction, and abundance determination computationally feasible. We ran various trials and found our results to be insensitive to bin widths.

\begin{figure*}
\psfrag{O}[c][][1.]{$\odot$}
\psfrag{S}[c][][1.]{$\star$}
\centering{\includegraphics[scale=1.,width=0.45\textwidth,trim=0.pt 0.pt 190.pt 70.pt,clip]{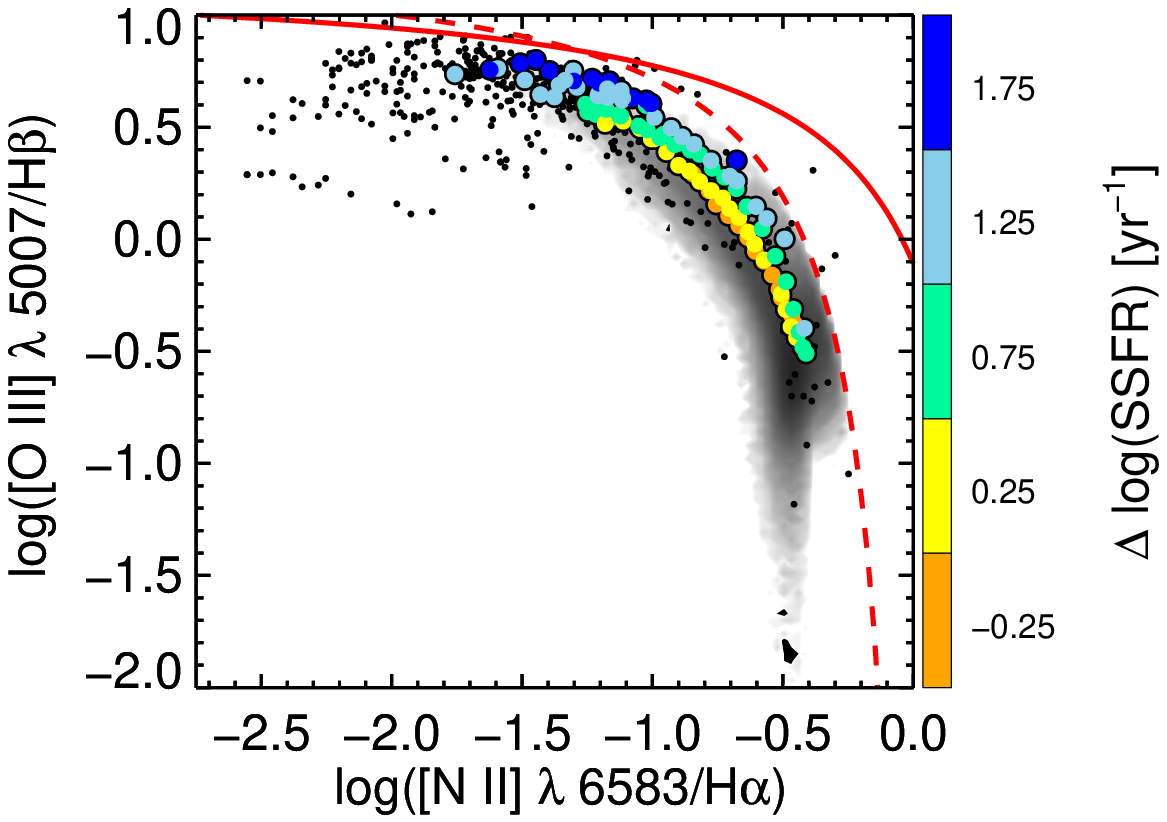}}
\centering{\includegraphics[scale=1.,width=0.45\textwidth,trim=0.pt 0.pt 190.pt 70.pt,clip]{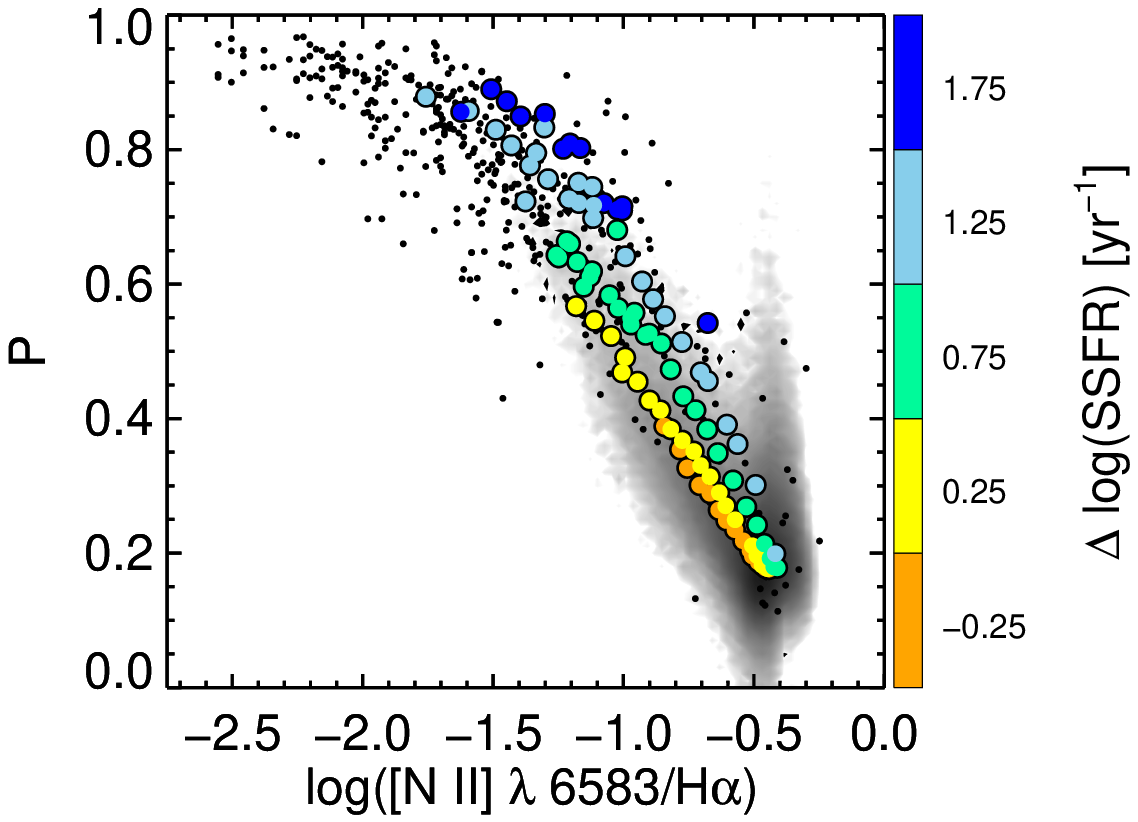}}
\caption{Left: BPT diagram of the stacks (circles) relative to SDSS star forming galaxies (gray contours) and \ion{H}{ii} regions from \citet{Pilyugin12} (small black points). Color coding is done according to $\dssfr$. The dashed and solid red lines are from \citet{Kauffmann03b} and \citet{Kewley06} respectively and denote the boundaries between star forming galaxies and AGN. Right: Excitation parameter $P \equiv R_3/\rtwothree$ versus $\log(N2)$. The symbol notation is the same as the left panel. In the high excitation regime, the stacks and SDSS galaxies closely resemble \ion{H}{ii} regions. At lower excitation (where the majority of SDSS galaxies are located) there are very few \ion{H}{ii} regions; the two populations are clearly subject to different conditions.}
\label{fig:BPT}
\end{figure*}

The left panel of Figure~\ref{fig:BPT} shows where $\mstar$--$\dssfr$ stacks fall on the BPT diagram relative to the galaxies in our sample (gray contours) and individual \ion{H}{ii} regions from \citet{Pilyugin12} (black points). The stacks with high $\dssfr$ are undergoing relatively intense star formation, and their line ratios closely resemble those of individual \ion{H}{ii} regions. The passively star forming stacks track the overall distribution of galaxies, which is not traced by the individual \ion{H}{ii} regions.

Naively we expect that galaxies undergoing more intense star formation have many more ionizing photons per atom. While the excitation parameter $P$ is marginally dependant on abundance, the right panel of Figure~\ref{fig:BPT} suggests our naive expectation is correct; stacks with high $\dssfr$ show systematically higher values of $P$. Incorporating $\dssfr$ accounts for some of the strong line ratios' sensitivity to ionization conditions.  

Lastly, it is easily shown that many strong line ratios (e.g. N2) are biased by SFR since they include $\halpha$ flux. By grouping galaxies with similar $\dssfr$, which is equivalent to SFR at fixed $\mstar$, our chosen stacking methodology minimizes this bias. 

\subsection{Stellar Continuum Subtraction}
Many emission lines used in this study (particularly \oaur) fall in wavelength regimes where stellar absorption features are present. Therefore it is necessary to fit and remove the underlying stellar population contribution to the stacked spectra. Following \citet{Andrews13}, we use the STARLIGHT spectral synthesis code \citep{CidFernandes05, CidFernandes11} and a library of 300 empirical \verb|MILES| spectral templates \citep{SanchezBlazquez06, Cenarro07,Vazdekis10,FalconBarroso11} to generate a synthetic spectrum representative of the underlying stellar population for each of our stacks. We adopt the \citet{Cardelli89} extinction law and mask the locations of all bright emission lines.

For strong lines redward of 4000\AA\ (\hbeta, [\ion{O}{iii}]~$\lambda \lambda$4959, 5007\AA, \halpha, [\ion{N}{ii}]~$\lambda \lambda$6548, 6583\AA, and [\ion{S}{ii}]~$\lambda \lambda$6716, 6731\AA)  we model the stellar continuum using template fits to the entire spectral range (3700\AA--7360\AA). We fit the continuum near weaker emission lines, auroral lines, and strong lines blueward of 4000\AA\ using template fits to the continuum within a few 100\AA\ of each line since this provides a significant reduction in the rms of the continuum around the line \citep{Andrews13}. See Table~\ref{tab:wave} for details regarding each emission line's fit region.

\begin{table}
\caption{Wavelength Fit and Mask Ranges of Measured Lines. \label{tab:wave}}
\begin{minipage}{140mm}
\begin{tabular}{@{}lcc}
\hline
\hline
{Line$^a$} & {Fit Range$^b$} & {Mask Range$^c$} \\
\hline
{}[O~{\sc ii}]~$\lambda$3727  &  3700--4300  &  3710--3744 \\
{}[Ne~{\sc iii}]~$\lambda$3868  &  3800--4100  &  3863--3873 \\
{}[S~{\sc ii}]~$\lambda$4069  &  3950--4150  &  \nodata \\
{}H$\gamma$~$\lambda$4340  &  4250--4450  &  4336--4344 \\
{}[O~{\sc iii}]~$\lambda$4363  &  4250--4450  &  4360--4366 \\
{}He~{\sc ii}~$\lambda$4686  &  4600--4800  &  4680--4692 \\
{}[Ar~{\sc iv}]~$\lambda$4740  &  3700--7360  &  \nodata \\
{}H$\beta$~$\lambda$4861  &  3700--7360  &  4857--4870 \\
{}[O~{\sc iii}]~$\lambda$4959  &  3700--7360  &  4954--4964 \\
{}[O~{\sc iii}]~$\lambda$5007  &  3700--7360  &  5001--5013 \\
{}[N~{\sc ii}]~$\lambda$5755  &  5650--5850  &  5753--5757 \\
{}[S~{\sc iii}]~$\lambda$6312  &  6100--6500  &  6265--6322 \\
{}[N~{\sc ii}]~$\lambda$6548  &  3700--7360  &  6528--6608 \\
{}H$\alpha$~$\lambda$6563  &  3700--7360  &  6528--6608 \\
{}[N~{\sc ii}]~$\lambda$6583  &  3700--7360  &  6528--6608 \\
{}[S~{\sc ii}]~$\lambda$6716  &  3700--7360  &  6696--6752 \\
{}[S~{\sc ii}]~$\lambda$6731  &  3700--7360  &  6696--6752 \\
{}[Ar~{\sc iii}]~$\lambda$7135  &  7035--7235  &  7130--7140 \\
{}[O~{\sc ii}]~$\lambda$7320  &  7160--7360  &  7318--7322 \\
{}[O~{\sc ii}]~$\lambda$7330  &  7160--7360  &  7328--7332 \\

\hline
\hline
\end{tabular}

\medskip
$^a$Emission lines.

$^b$The wavelength range of the stellar continuum fit.

$^c$The wavelength range of the stellar continuum fit \\ 
that was masked out.

\end{minipage}
\end{table}

\subsection{Line Flux Measurement}
\label{sec:line_flux}
Following \citet{Andrews13}, we fit the emission lines of the stacked spectra using the \textit{specfit} routine \citep{Kriss94} in the \verb|IRAF/STSDAS| package. We use the simplex $\chi^2$ minimization algorithm to simultaneously fit a flat continuum and Gaussian profile to each emission line. \citet{Andrews13} found this to be a robust method consistent with other flux measurement techniques. Uncertainties are derived from the $\chi^2$ of the fit returned by \textit{specfit}. We deredden the spectra using the extinction law from \citet{Cardelli89} and the assumption of case B recombination (\halpha/\hbeta = 2.86 for $T_e = 10^4$ K). \citet{Andrews13} estimate the systematic error introduced by adopting a fixed \halpha/\hbeta\ ratio to be $\lesssim 0.07$ dex. Finally, with the exception of [\ion{O}{ii}]~$\lambda 3727$ \AA, our diagnostic emission lines are anchored to nearby Balmer lines, and are thus insensitive to reddening.

\section{Analysis}
\label{sec:analysis}

\subsection{Abundances}
\label{sec:abund}

\begin{table}
\caption{Line Fluxes \label{table:flux}}
\begin{minipage}{140mm}
\begin{tabular}{@{}lcl}
\hline
\hline
{Column} & {Format} & {Description} \\
\hline
1  &   F5.2   &  Lower stellar mass limit of the stack \\
2  &   F5.2   &  Upper stellar mass limit of the stack \\
3  &   F5.2   &  Lower SFR limit of the stack \\
4  &   F5.2   &  Upper SFR limit of the stack \\
5  &   I5     &  Number of galaxies in the stack \\
6  &   F8.3   &  Oxygen abundance of the stack \\
7  &   F8.3   &  Error on oxygen abundance \\
8  &   F8.3   &  [\ion{O}{ii}] $\lambda$3726 line flux \\
9  &   F8.3   &  Error on [\ion{O}{2}] $\lambda$3726 line flux \\
10  &  F8.3   &  [\ion{O}{ii}] $\lambda$3729 line flux \\

\hline
\hline

\end{tabular}

\medskip

(This table is published in its entirety in the electronic edition \\
of the journal. The column names are shown here for guidance \\
regarding its form and content.)

\end{minipage}
\end{table}

We compute the chemical abundances of the stacks using the same procedure as \citet{Andrews13}; here we present a brief overview and direct the reader to that paper for further details.

We assume a simple two-zone model composed of a high ionization region (traced by [\ion{O}{iii}]) and a low ionization region (traced by [\ion{O}{ii}], [\ion{N}{ii}], and [\ion{S}{ii}]). Previous works have assumed simple relationships between the temperatures of the high and low ionization regions \citep[the $\ttwothree$ relation][]{Campbell86, Garnett92, Pagel92, Izotov06, Pilyugin09}. We assume a linear $\ttwothree$ relation normalized such that we get the best agreement in stacks for which we are able to measure the temperature of both ionization zones (see below). We use a Monte Carlo technique to derive uncertainties in our measurements.

We measure the electron temperature and density using the \verb|IRAF/STSDAS| \textit{nebular.temden} routine \citep{Shaw95}, which is based on the 5 level atom from \citet{DeRobertis87}. We use the $\sulratio$ ratio to measure the electron density. We use the auroral oxygen ratios ($\oaurratiottwo$ and $\oaurratiotthree$) to measure $T_2$ and $T_3$ respectively. \citet{Andrews13} discuss at length the differences between the canonical $\ttwothree$ relation and that observed for their stacks and find that in general their stacks fall below the \citet{Campbell86} relation (in the sense of low $T_2$ at fixed $T_3$). This offset from the predicted relation has been previously seen \citep{Pilyugin10}. The fact that this offset disappears for galaxies with relatively high SFRs (which are likely to have contributions from relatively young stellar populations) indicates that the offset is likely due to the differences between the single stellar spectra used by \citet{Stasinska82} and the composite \ion{H}{ii} region spectrum that ionizes the gas in a galaxy.

The ionic abundances of O$^+$ and O$^{++}$ are calculated using the electron temperature, electron density, the flux ratios of the strong lines relative to $\hbeta$, and the \verb|IRAF/STSDAS| \textit{nebular.ionic} routine \citep{DeRobertis87,Shaw95}. Atomic data plays a critical role in direct method temperature determinations \citep{Kennicutt03}. For example, \citet{Berg15} noted a substantial difference in \ion{S}{iii} temperatures when using updated collision strengths. The \ion{O}{iii} temperatures are largely unaffected by the updated atomic data, so we utilize the \textit{nebular.temden} routine without modification. The uncertainties in the abundances of individual ionic species are determined with the same Monte Carlo simulations used to determine the uncertainties in electron temperatures. The ionic abundance uncertainties are used to analytically calculate the uncertainty in the total abundances.

We assume the total oxygen abundance is given by
\begin{equation}
\frac{\rm O}{\rm H} = \frac{\rm O^+}{\rm H^+} + \frac{\rm O^{++}}{\rm H^+}.
\end{equation}
Historically, the temperature of the high ionization region, $T_3$, is measured using the direct method and $T_2$ is then inferred using the $\ttwothree$ relation. At high masses, we are unable to measure $T_3$ but often have a measurement of $T_2$. We use the stacks where both $T_2$ and $T_3$ are measured to infer a $\ttwothree$ relation that results in the best agreement between measured and inferred $T_3$. As in \citet{Andrews13}, this is done using a systematic shift ($\sim$0.1 dex) in the $\log$(O/H) of the stacks for which $T_2$ was measured and used to infer $T_3$.

\subsection{Empirical Calibrations}
There are many abundance diagnostic ratios. Our choice of ratios to consider is motivated by three factors: (1) our calibration(s) should be empirical, (2) the distribution of line ratios for individual galaxies in a stack ought to be reasonably peaked around the mean value, and (3) the calibration ought to be valid for the majority of our stacks. 

The most commonly used oxygen abundance diagnostics are N2 and O3N2 \citep{Denicolo02,Pettini04,Marino13}, N2O2 \citep{Dopita00,Kewley02}, and R23 \citep{Pagel79,McGaugh91,Pilyugin03,Kobulnicky04} . Figure~\ref{fig:dssfr} shows the distribution of individual galaxies in the $\mstar$--diagnostic planes for these diagnostics. In panel ``(a)'' of Figure~\ref{fig:dssfr}, the distribution of galaxies is such that galaxies with similar $\mstar$ and $\dssfr$ follow a relatively tight sequence in the $\mstar$--N2 plane. Similar behavior is seen in panel ``(b)'' (O3N2) and, to a somewhat lesser extent, panel ``(c)'' (N2O2). In panel ``(d)'' at fixed $\mstar$ and $\dssfr$, the values of $\rtwothree$ follow a relatively broad distribution; the scatter in $\rtwothree$ at fixed $\mstar$ and $\dssfr$ can be comparable to the entire range spanned by the diagnostic. In this instance, the degree to which the average strong line value of a given stack is representative of the galaxies within that stack is less meaningful than with other diagnostics. This is a primary concern when stacking galaxies (see Footnote 14 of \citet{Salim14} for an example of how binning can lead to the wrong impression).

An additional concern with strong line abundance diagnostics is the effect of ionization parameter variations on the diagnostic ratios \citep{Kewley02,Steidel14}. The ionization parameter $\Gamma$ is given by
\begin{equation}
\Gamma \equiv \frac{\Phi}{n_H} \approx \frac{\Phi}{n_e}
\end{equation}
where $n_H$ is the number density of hydrogen atoms and $\Phi$ is the density of hydrogen ionizing photons. Changes in the ionization parameter can be due to either variations in the temperature of the ionizing continuum (i.e. a galaxy composed of systematically hotter stars than average) and/or variations in the physical conditions of star forming regions (i.e. higher stellar densities and/or lower gas densities than average). In order to eliminate these biases, it would be advantageous to use a diagnostic that is insensitive to ionization parameter variations \citep[e.g. N2O2,][]{Kewley02}, though our choice of $\dssfr$ as a second parameter should at least somewhat account for differences in ionization parameter (see the right panel of Figure~\ref{fig:BPT}). 

The N2 diagnostic is subject to biases caused by the ionization parameter as well as the hardness of the ionization spectrum \citep{Kewley02}, but has been shown to be a useful abundance diagnostic in high excitation regions \citep{TSB94,Binette96,Pettini04, Marino13}. Furthermore, [\ion{N}{ii}] $\lambda$ 6583 and $\halpha$ are closely spaced, making their ratio insensitive to variations in reddening corrections. The O3N2 diagnostic is also sensitive to ionization parameter \citep{Kewley02}, but is less sensitive to variations in the hardness of the ionizing spectrum than N2 \citep{Kewley13a, Brown14, Steidel14}. N2O2 is insensitive to ionization parameter, but is dependent on the secondary nature of nitrogen \citep{Kewley02}. We will use N2O2 to estimate the effect of ionization parameter variations on the other diagnostics.

With the above considerations in mind, we focus the remainder of our analysis on the N2, O3N2, and N2O2 strong line diagnostics.  As discussed above, the distribution of $\rtwothree$ at fixed $\mstar$ and $\dssfr$ is not strongly peaked. Furthermore, the double valued nature of $\rtwothree$ requires that an additional diagnostic sensitive to ionization parameter be used in conjunction with an iterative method to solve for an oxygen abundance. This precludes the empirical nature of our calibrations. Most importantly, a large fraction of our galaxies fall within the ``transition zone'' of the $\rtwothree$ diagnostic, where the diagnostic is insensitive to oxygen abundance \citep{Dopita13}. As a result, we refrain from further consideration of $\rtwothree$. 

\section{Results}
\label{sec:results}

In Section~\ref{sec:justify_stack} we demonstrated with Figure~\ref{fig:dssfr} that each $\mstar$-$\dssfr$ stack has characteristic diagnostic line ratios which are representative of the individual galaxies in that stack. Following previous works \citep[e.g.,][]{Alloin79, Pettini04, Marino13} we combine these diagnostic ratios with direct method oxygen abundances to derive a relationship between the two. \citet{Salim14} showed that at fixed $\mstar$ we expect galaxies with low (high) $\dssfr$ to be offset from the star forming main sequence in the sense of high (low) oxygen abundance. Given the strong correlation between our diagnostic ratios and $\mstar$, we assume the following form for our empirical calibrations:
\begin{equation}
12+\log({\rm O/H}) = f_1(X) + f_2(\dssfr)
\end{equation}
where $X$ is a particular diagnostic value (e.g. N2) and $f_1$ and $f_2$ are functions of the respective variables. For simplicity, we assume $f_1$ and $f_2$ are each linear functions in their respective parameter, except for the case of N2 where we allow $f_1$ to take the form of a second degree polynomial. We use \verb|MPFIT| \citep{Markwardt09}, an IDL implementation of the robust non-linear least square fitting routine \verb|MINPACK-1|, to fit the relationship between $\log$(O/H), $X$, and $\dssfr$.

From Equation~\ref{eq:dssfr}, it is clear that for a galaxy with a known $\mstar$ and SFR, $\dssfr$ then only depends on the average SSFR at that $\mstar$. In practice, we compute the median SSFR in $\mstar$ bins 0.1 dex wide. However, a good approximation for $\langle \log$(SSFR)$\rangle_{\small{\mstar}}$ as a function of $\mstar$ is:

\begin{multline}
\langle \log({\rm SSFR})\rangle_{\small{M_{\star}}} = 283.728 - 116.265 \times \log\mstar + \\ 17.4403 \times \log\mstar^2 - 1.17146 \times \log\mstar^3 + 0.0296526 \times \log\mstar^4.
\label{eq:sfms}
\end{multline}

\noindent
We provide this form rather than the expression from \citet{Salim07} because the two begin to diverge below $\log(\mstar/\msol) \sim 9$.

\subsection{N2 Method}
\label{sec:N2}

Our new calibration of $\ohte$ based on N2 and $\dssfr$ is

\begin{multline}
12 + \log({\rm O/H})_{\rm N2} = 9.12 + 0.58 \times \log({\rm N2}) \\
   - 0.19 \times {\dssfr}.
\end{multline}

\begin{figure}
\psfrag{O}[c][][1.]{$\odot$}
\psfrag{S}[c][][1.]{$\star$}
\centering{\includegraphics[scale=1.,width=0.5\textwidth,trim=0.pt 0.pt 190.pt 0.pt,clip]{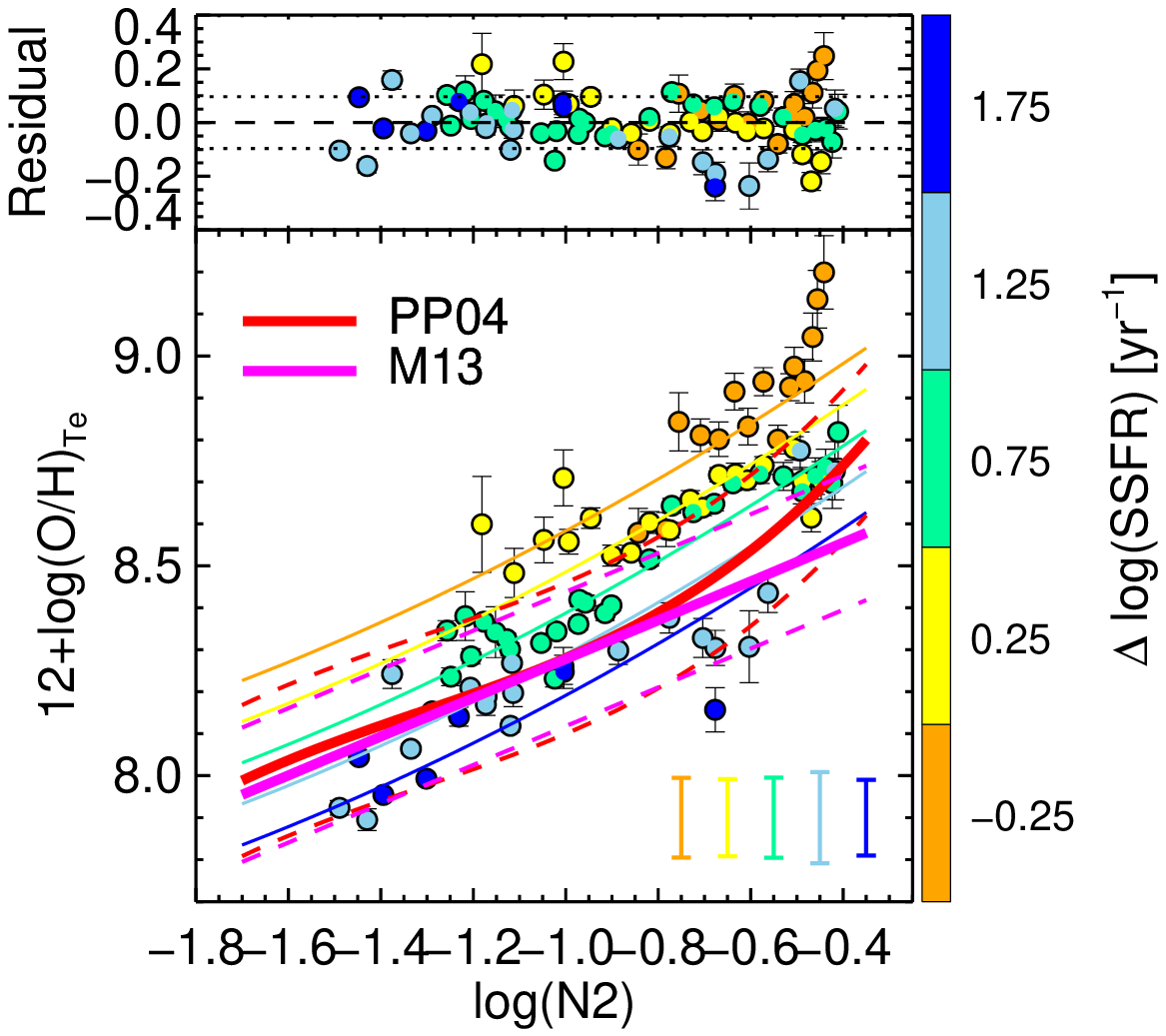}}
\caption{Direct method oxygen abundances of the stacks as a function of N2 and $\dssfr$. The circles show the actual measurements; the various lines show our fit to $\ohte$ as a function of N2 and $\dssfr$. The thick red and magenta lines shows the fits from \citet{Pettini04} and \citet{Marino13}, respectively, which are based almost entirely on direct method oxygen abundances of individual \ion{H}{ii} regions. The top panel shows the residuals of the fit; the dashed lines show the RMS of the residuals.} 
\label{fig:N2}
\end{figure}

Figure~\ref{fig:N2} shows that the slope of the relationship between $\ohte$ and N2 at fixed $\dssfr$ is comparable to the slope of \citet{Pettini04} (red line) and \citet{Marino13} (magenta line), and agree well for the galaxies with high $\dssfr$. This agrees with previous studies \citep[e.g.][]{Brown14} which have shown that those empirical relations accurately predict $\ohte$ for high excitation galaxies. This is not particularly surprising because galaxies with very compact, high star formation rates for a given \mstar\ are similar to individual \ion{H}{ii} regions in terms of excitation conditions.

As one moves from high excitation galaxies toward the star forming main sequence, the population of galaxies tends toward lower excitation conditions than the \ion{H}{ii} regions used in \citet{Pettini04}. The observational consequence is that SDSS galaxies have higher \ohte than predicted by previous calibrations at a given value of N2.

For galaxies above $\sim Z_{\odot}$, N2 saturates as it becomes the dominant coolant of the ISM \citep{Baldwin81,Pettini04}. This explains the pile up of stacks around $\log$(N2)$\approx -0.5$ in Figure~\ref{fig:N2} for the low $\dssfr$ stacks. As a result this calibration becomes unreliable when the line ratio reaches this value. The top panel of Figure~\ref{fig:N2} shows the residuals of the fit. It is clear that the quality of the calibration worsens at high metallicities. We include the RMS of the residuals in Table~\ref{tab:fits}.

\subsection{O3N2 Method}
\label{sec:O3N2}

Our new calibration of $\ohte$ based on O3N2 and $\dssfr$ is

\begin{multline}
12 + \log({\rm O/H})_{\rm O3N2} = 8.98 - 0.32 \times \log({\rm O3N2}) \\
   - 0.18 \times {\dssfr}.
\end{multline}

\begin{figure}
\psfrag{O}[c][][1.]{$\odot$}
\psfrag{S}[c][][1.]{$\star$}
\centering{\includegraphics[scale=1.,width=0.5\textwidth,trim=0.pt 0.pt 190.pt 0.pt,clip]{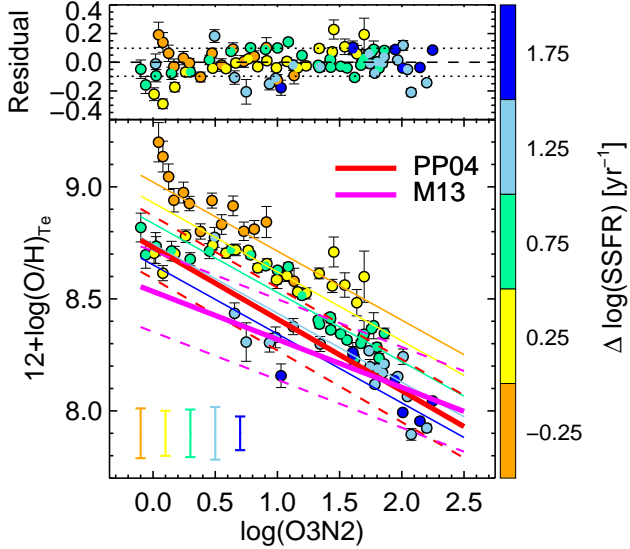}}
\caption{Same as Figure~\ref{fig:N2} but for the O3N2 diagnostic.} 
\label{fig:O3N2}
\end{figure}

Figure~\ref{fig:O3N2} shows that the slope of the relationship between $\ohte$ and O3N2 at fixed $\dssfr$ is comparable to the slope of \citet{Pettini04} (thick red line) and \citet{Marino13} (thick magenta line), and agrees well for the galaxies with high $\dssfr$. Again this is in agreement with \citet{Brown14}, who showed that high excitation galaxies with significant populations of young stars are essentially indistinguishable from individual \ion{H}{ii} regions from the perspective of a diagnostic ratios. We do find a marginally steeper slope than \citet{Marino13}. This could be due to a selection effect because at high (low) metallicities we lack high (low) $\dssfr$ bins, which could artificially steepen our calibration.  In addition, the steepness of the \citet{Pettini04} calibration may be due to the photoionization models used at high metallicities. The \citet{Marino13} calibration suffers no such bias, since their measurements are based entirely on individual Hii regions. More data are needed to explore this possibility further.

Closer to the star forming galaxy main sequence, the calibration presented here begins to diverge from the previous calibrations based on \ion{H}{ii} regions. Again, this is because the galaxies on the star forming main sequence display lower excitation conditions than the \ion{H}{ii} regions used in the previous calibrations.

The O3N2 diagnostic performs better than the N2 diagnostic at high $\ohte$. While N2 saturates at high metallicity, the intensity of collisionally excited oxygen lines is still falling with increasing oxygen abundance.

\subsection{N2O2 Method}
\label{sec:N2O2}

Our new calibration of $\ohte$ based on N2O2 and $\dssfr$ is

\begin{multline}
12 + \log({\rm O/H})_{\rm N2O2} = 9.20 + 0.54 \times \log({\rm N2O2}) \\
   - 0.36 \times {\dssfr}.
\end{multline}

\begin{figure}
\psfrag{O}[c][][1.]{$\odot$}
\psfrag{S}[c][][1.]{$\star$}
\centering{\includegraphics[scale=1.,width=0.5\textwidth,trim=0.pt 0.pt 190.pt 0.pt,clip]{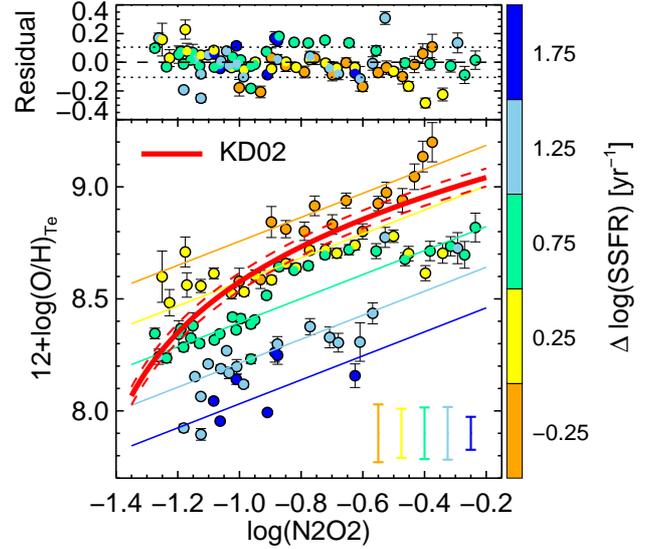}}
\caption{Same as Figure~\ref{fig:N2} but for the N2O2 diagnostic. The thick red line shows the fit from \citet{Kewley02}.} 
\label{fig:N2O2}
\end{figure}

In Figure~\ref{fig:N2O2} we compare our measurements from the stacks with the N2O2 calibration from \citet{Kewley02}. At high metallicities, we find excellent agreement between the star forming galaxy main sequence of our stacks and the calibration from \citet{Kewley02}. This could be due to the fact that this calibration is insensitive to ionzation parameter. At fixed N2O2, stacks with high $\dssfr$ show lower $\ohte$ than stacks with lower $\dssfr$, as one would expect in the case of inflow driven star formation.

\citet{Kewley02} explicitly state that the N2O2 calibration should only be used above $12+\log$(O/H)$ > 8.6$ since this diagnostic derives its utility from the secondary nature of nitrogen at high metallicity. However, in the context of galaxy evolution where inflows and outflows have a strong effect on the oxygen abundance we argue that this selection criteria should instead be based on the value of the N2O2 diagnostic itself. For instance, consider a galaxy which has undergone prolonged star formation and enriched its ISM well above solar metallicity such that the secondary nature of nitrogen is unambiguous. Now, suppose this galaxy were to accrete a substantial amount of gas from the IGM. The ISM would be diluted, the metallicity would decrease, and the SFR would increase. The galaxy would move off the main sequence, increasing $\dssfr$. All the while, the N2O2 ratio would remain largely unchanged, since the relative abundance of nitrogen and oxygen is unaffected by inflows of pristine gas \citep{Koppen05,Masters14}. The high SFR stacks shown in Figure 14 of \citet{Andrews13} are consistent with this picture of inflow driven dilution. Nitrogen can be secondary even at low metallicities, provided the galaxy is sufficiently chemically evolved.

Figure 3 of \citet{Kewley02} shows that the N2O2 diagnostic becomes sensitive to metallicity at $\log$(N2O2) $\sim -1.25$. Our Figure~\ref{fig:N2O2} illustrates that this happens at the lower range probed by our stacks. The $\ohte$ of our stacks does show a clear dependence on N2O2, even at low metallicities. Unevolved galaxies for which nitrogen is still primary could potentially contaminate the stacks. However, the left panel of Figure~\ref{fig:dssfr} shows that there are relatively few galaxies with $\log$(N2O2) $< -1.25$. Thus we are confident our N2O2 calibrations are valid even though we apply them at low metallicities. 

\subsection{Which Calibration Is Best?}
\label{sec:use}

\begin{table*}
\caption{Calibration results. \label{tab:fits}}
\begin{minipage}{140mm}
\begin{tabular}{@{}lccccc}
\hline
\hline
{Diagnostic} & {$a$} & {$b$} & {$c$} & {$d$} & {rms Residuals} \\
\hline
N2\ldots\ldots\ldots\ldots\ldots\ldots\ldots    & 9.25  & 0.83   & 0.12     &  -0.20 &   0.0965  \\
O3N2\ldots\ldots\ldots\ldots\ldots\ldots        & 8.98  & -0.32  & \nodata  &  -0.18 &   0.0976  \\
N2O2\ldots\ldots\ldots\ldots\ldots\ldots        & 9.20  & 0.54   & \nodata  &  -0.36 &   0.1053  \\
\hline
\multicolumn{6}{c}{Star Forming Main Sequence} \\
\hline
\multicolumn{6}{c}{$\langle \log({\rm SSFR})\rangle_{\small{M_{\star}}} = 283.728 - 116.265 \times \log\mstar + 17.4403 \times \log\mstar^2 - 1.17146 \times \log\mstar^3 + 0.0296526 \times \log\mstar^4$} \\

\hline
\hline
\end{tabular}

\medskip

\end{minipage}
\end{table*}

\begin{figure*}
\psfrag{O}[c][][1.]{$\odot$}
\psfrag{S}[c][][1.]{$\star$}
\centering{\includegraphics[scale=1.,width=\textwidth,trim=0.pt 35.pt 0.pt 160.pt,clip]{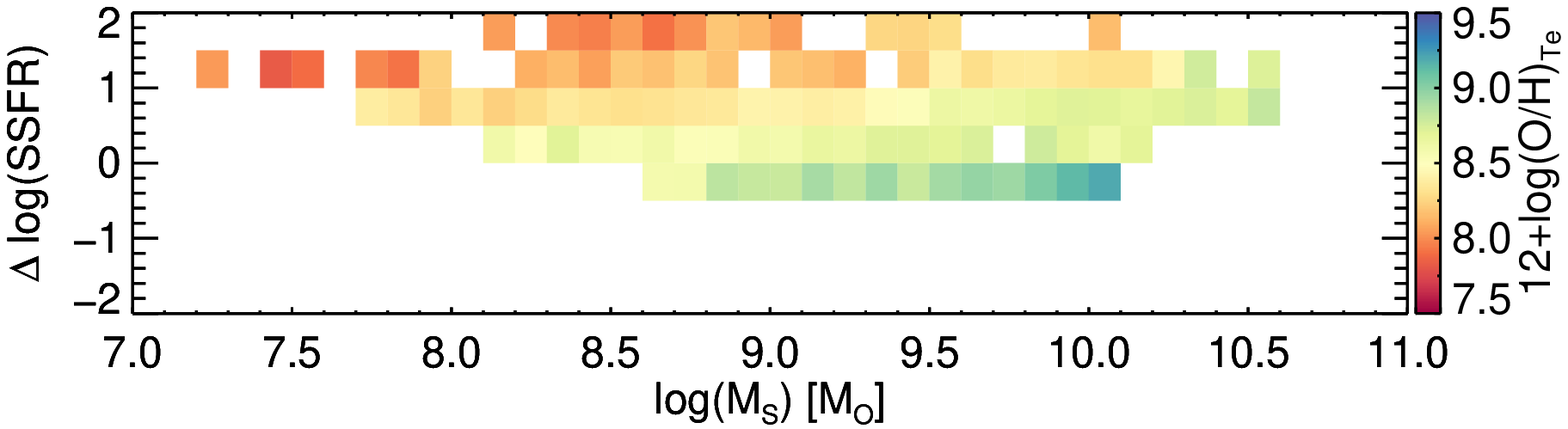}}
\centering{\includegraphics[scale=1.,width=\textwidth,trim=0.pt 35.pt 0.pt 160.pt,clip]{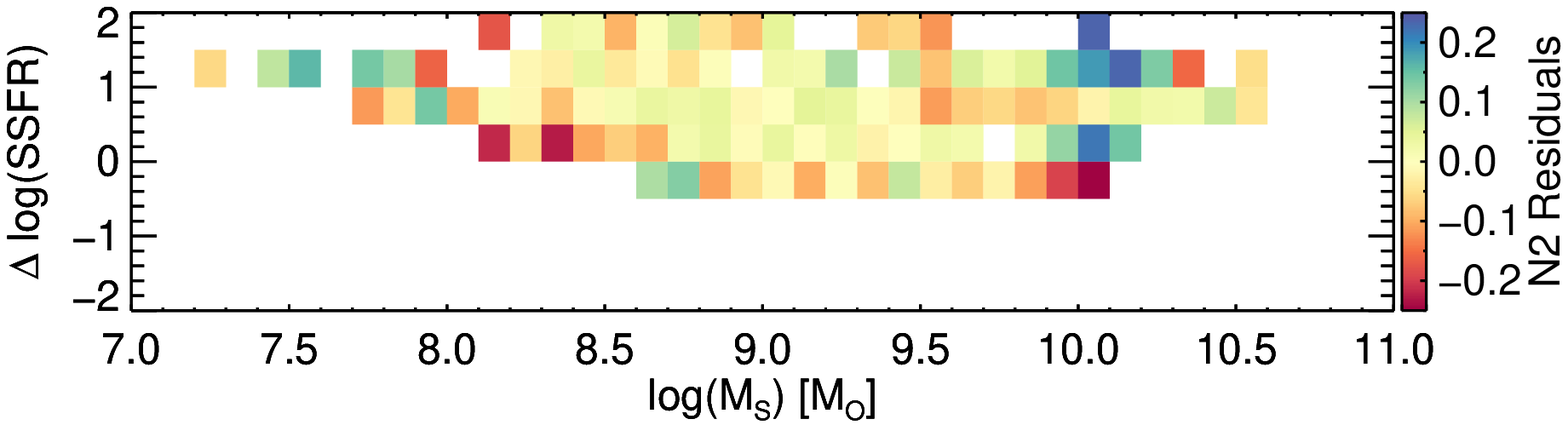}}
\centering{\includegraphics[scale=1.,width=\textwidth,trim=0.pt 35.pt 0.pt 160.pt,clip]{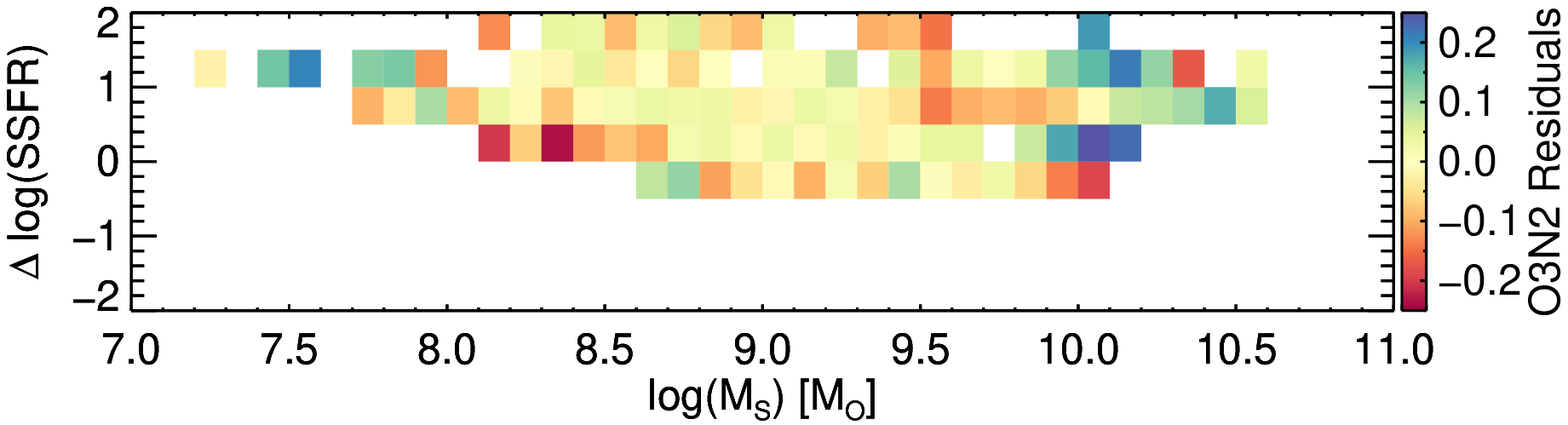}}
\centering{\includegraphics[scale=1.,width=\textwidth,trim=0.pt 0.pt 0.pt 160.pt,clip]{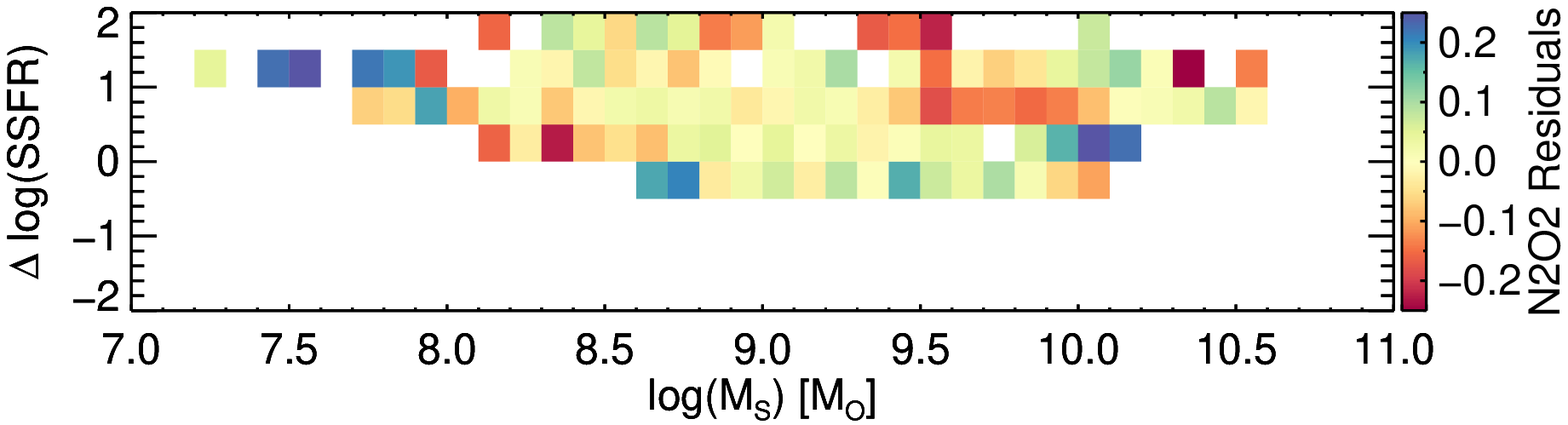}}
\caption{Overview of our binning and quality of our strong-line calibrations. The top panel shows the distribution of direct method measurements in $\mstar$-$\dssfr$ space. Each square represents a $\mstar$-$\dssfr$ stack. The color coding denotes metallicity. Metallicity generally increases as $\mstar$ increases and/or $\dssfr$ decreases. The second, third, and fourth panels show the residuals for the N2, O3N2, and N2O2 diagnostics respectively. All three diagnostics perform well across most of the parameter space. The O3N2 diagnostic was the most accurate (47/110 bins), followed by N2O2 (33/110 bins) and N2 (30/110 bins).} 
\label{fig:grids}
\end{figure*}

Figure~\ref{fig:grids} summarizes our results in $\mstar$--$\dssfr$ space and illustrates several systematic effects correlated with $\mstar$ and/or $\dssfr$.

The top panel shows the distribution of stacks with measured $\ohte$ in $\mstar$-$\dssfr$ space. The color of each square reflects the metallicity. The second, third, and fourth panels show the residuals for the N2, O3N2, and N2O2 diagnostics, respectively. Red indicates where the strong line diagnostic overestimates the direct method metallicity, while blue indicates the alternative. Column $d$ in Table~\ref{tab:fits} shows the mean residuals for each calibration. On average the calibrations are accurate to within 0.10 dex, although there are typically 2-3 stacks for each diagnostic that have substantially larger residuals. The calibrations perform worse for the highest metallicity stacks. This is evident in residuals shown in the top panels of Figures~\ref{fig:N2}, \ref{fig:O3N2}, and~\ref{fig:N2O2}. The metallicities of the lowest mass stacks are also difficult to accurately predict. This is likely due to the small number ($\sim 5$) of galaxies in these stacks. One or two galaxies with anomoulous line ratios can significantly influence the line ratios of the stack \citep{Andrews13}.

In general, no single calibration vastly outperforms the others, though O3N2 does fare slightly better. O3N2 was the preferred diagnostic for 43\% (47/110) of the stacks, followed by N2O2 with 30\% (33/110), and N2 was ranked last with 27\% (30/110). There does not appear to be any systematic trend where one calibration does better than the others, though N2O2 is only marginally worse than O3N2 for many of the stacks and is subject to fewer biases.

The N2O2 calibration has a larger dependence on $\dssfr$ (0.36, column $d$ in Table~\ref{tab:fits}) than the other calibrations ($\sim 0.2$). This likely reflects the fact that N2 and O3N2 are sensitive to ionization parameter, whereas N2O2 is not. At fixed metallicity, a systematically high ionization parameter (correlated with high $\dssfr$) biases the N2 and O3N2 line ratios in the direction of low metallicity. Thus stacks with high $\dssfr$ have metal poor line ratios relative to a stack of lower $\dssfr$ and identical metallicity. This reduces the inferred dependence of metallicity on $\dssfr$. While all three calibrations perform equally well for our sample, these biases may be important considerations for applications to other samples. We emphasize that the rms residuals of the fit to the stacks does not reflect the actual precision of the calibration. As noted in Section~\ref{sec:justify_stack}, the reliability of the calibrations is primarily determined by the scatter in a given line ratio at fixed $\mstar$ and $\dssfr$, which is assumed to mean fixed O/H. This scatter is ultimately a function of $\mstar$, SFR, strong-line diagnostic, and sample selection. We include error bars in the lower corners of Figures~\ref{fig:N2}, \ref{fig:O3N2}, and~\ref{fig:N2O2} to show the typical uncertainty for our different $\dssfr$ bins, marginalized over $\mstar$. The error bars ($\sim 0.2$ dex) reflect the uncertainty in inferred O/H due to the scatter in strong-line ratio at fixed $\mstar$ and $\dssfr$, and typically exceed the widths of the O/H distributions in our bootstrap analysis ($\sim 0.15$ dex).

\begin{figure}
\psfrag{O}[c][][1.]{$\odot$}
\psfrag{S}[c][][1.]{$\star$}
\centering{\includegraphics[scale=1.,width=0.5\textwidth,trim=20.pt 0.pt 190.pt 70.pt,clip]{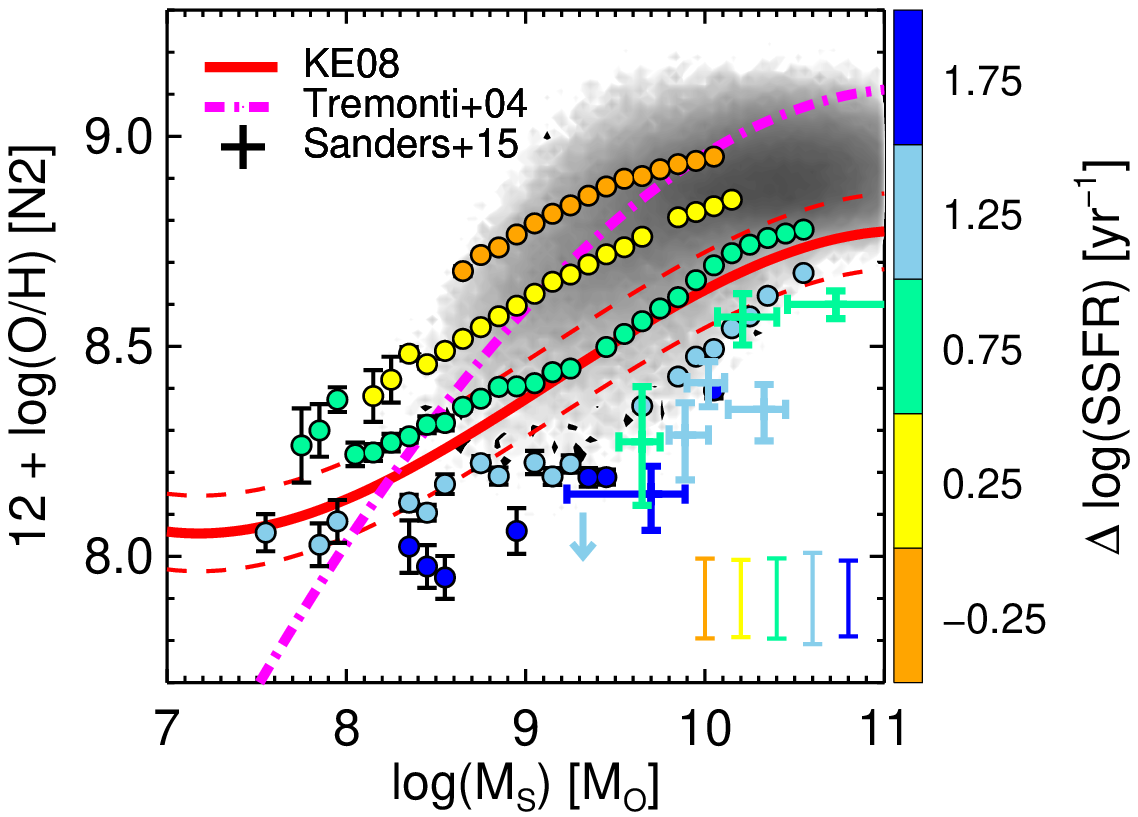}}
\caption{The MZR derived with our new N2 calibration. The circles represent our stacks, the crosses represent the high redshift star forming galaxy stacks from \citet{Sanders15}, and the gray contours represent the star forming SDSS galaxies used in our analysis. The thick red line shows the mass binned results from \citet{Kewley08}. The dot-dashed magenta line shows the MZR from \citet{Tremonti04}. The smooth behaviour of the stacks is ultimately the result of the average N2 varying so smoothly with $\mstar$. The galaxies from \citet{Sanders15} display lower (O/H) than the stacks with comparable $\mstar$ and $\dssfr$.} 
\label{fig:N2_MZR}
\end{figure}

\begin{figure}
\psfrag{O}[c][][1.]{$\odot$}
\psfrag{S}[c][][1.]{$\star$}
\centering{\includegraphics[scale=1.,width=0.5\textwidth,trim=20.pt 0.pt 190.pt 70.pt,clip]{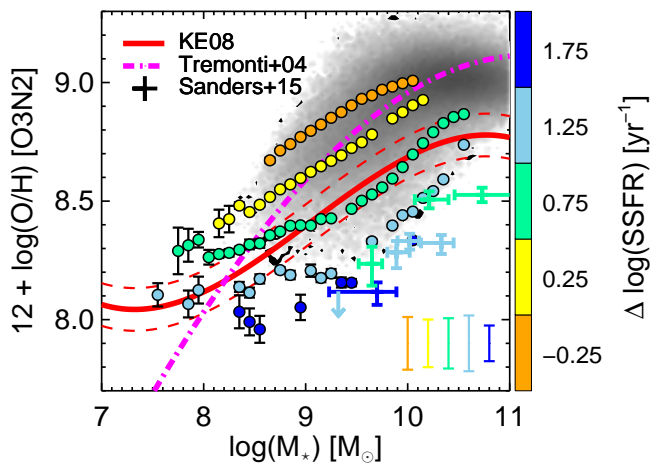}}
\caption{Same as Figure~\ref{fig:N2_MZR} but using the O3N2 diagnostic.}  
\label{fig:O3N2_MZR}
\end{figure}

\begin{figure}
\psfrag{O}[c][][1.]{$\odot$}
\psfrag{S}[c][][1.]{$\star$}
\centering{\includegraphics[scale=1.,width=0.5\textwidth,trim=20.pt 0.pt 190.pt 70.pt,clip]{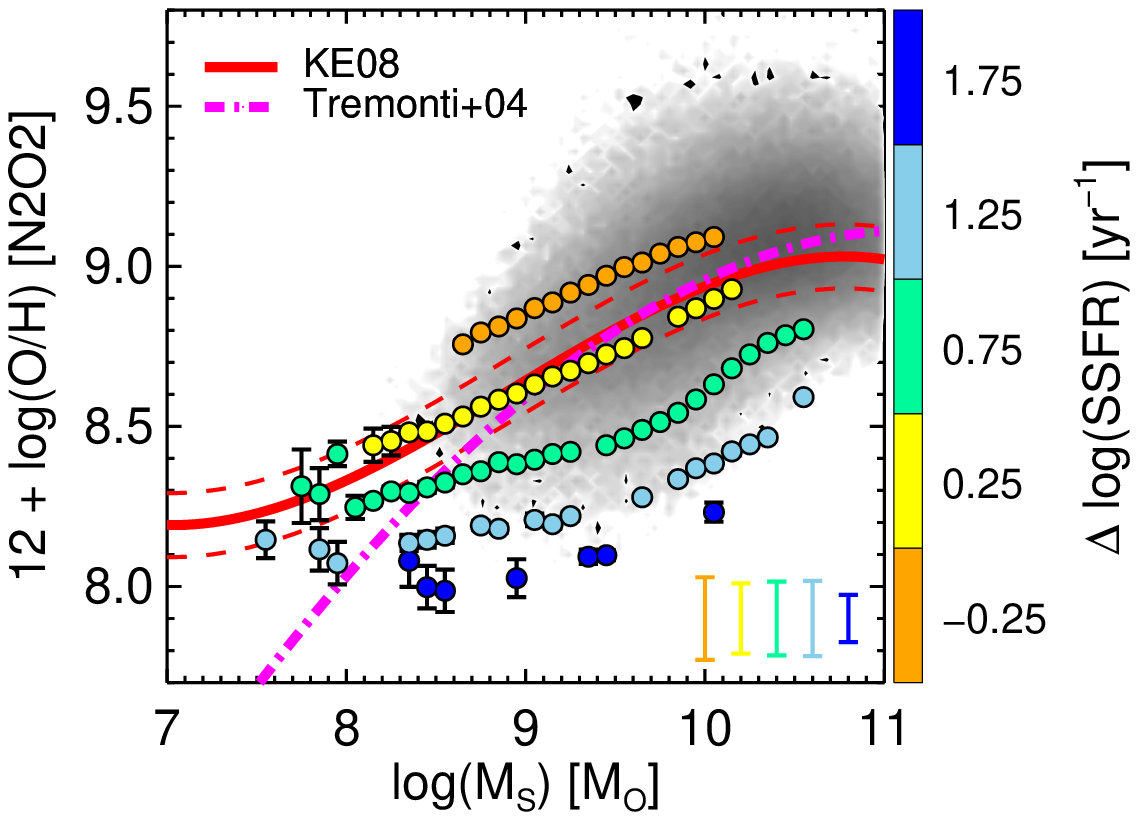}}
\caption{Same as Figure~\ref{fig:N2_MZR} but for the N2O2 diagnostic. The high redshift star forming galaxy stacks from \citet{Sanders15} are not included here, as the [\ion{O}{ii}]~$\lambda$3727 \AA\ line does not fall within the spectral range of the MOSFIRE instrument at $z \sim 2.3$. Note that the MZR resulting from this calibration has a higher normalization and larger scatter at fixed $\mstar$ than the other calibrations.} 
\label{fig:N2O2_MZR}
\end{figure}

\section{Discussion}
\label{sec:discussion}

\subsection{Application of New Calibrations to Local Galaxies}
\label{sec:local}

We first apply our newly derived strong line calibrations to the sample of individual star forming galaxies that went into our stacks. In Figures~\ref{fig:N2_MZR}, \ref{fig:O3N2_MZR}, and \ref{fig:N2O2_MZR} we show the distribution of SDSS galaxies (gray contours) and $\mstar$--$\dssfr$ stacks (colored points) in the M-Z plane. All metallicities are computed using the appropriate strong line calibration. In Figure~\ref{fig:N2_MZR} we apply the N2 calibration, in Figure~\ref{fig:O3N2_MZR} we apply the O3N2 calibration, and in Figure~\ref{fig:N2O2_MZR} we apply the new N2O2 calibration. In each panel the solid (dotted) red lines show the appropriate best fit MZR (scatter) from \citet{Kewley08}, in which the MZRs were measured by computing the median $\log$(O/H) as a function of mass. The dot-dashed magenta lines show the MZR from \citet{Tremonti04}.

If each $\mstar-\dssfr$ bin has a known $\ohte$, the uncertainty in the calibration is dominated by the average scatter in a given diagnostic at fixed $\ohte$. The error in any given measurement of $\ohte$ is typically much smaller than this. We estimate the scatter in a diagnostic at fixed $\ohte$ by averaging the scatter in the diagnostic over all masses at fixed $\dssfr$. These uncertainties are shown as error bars in the bottom corner of the plots and are generally comparable to the uncertainties in the calibrations ($\sim 0.10$ dex). The error bars on the points themselves represent the error on the mean. Due to the large number of galaxies in most stacks, the mean is typically measured to high precision.

In the case of N2 and O3N2, we find that our direct method strong line calibrations produce MZRs with higher (O/H) normalizations than the \citet{Kewley08} results, as expected from Figures~\ref{fig:N2} and~\ref{fig:O3N2}. In the case of N2O2, the normalization of the MZR is only marginally higher than the results from \citet{Kewley08}; this is due to the fact that, without accounting for $\dssfr$, our N2O2 calibration is very similar to that presented in \citet{Kewley02}. The slopes of all of our MZRs are roughly consistent with the results from \citet{Kewley08} and also appear to flatten at low masses ($\log(\mstar) \lesssim 8$). Each of the MZRs also agree well with the \citet{Tremonti04} MZR.

Figures~\ref{fig:N2} and~\ref{fig:O3N2} suggest that the $\dssfr_{1.0}^{1.5}$ bins should follow the \citet{Kewley08} MZR closest, when in fact it is the $\dssfr_{0.5}^{1.0}$ bins. This is purely a selection effect due to the difference in binning schemes. \citet{Kewley08} effectively binned in $\mstar$, whereas we have binned in both $\mstar$ and $\dssfr$. As shown in Figure~\ref{fig:dssfr} (top left), the $\mstar$--$\dssfr$ stacks with high $\dssfr$ have lower values of N2 than a corresponding mass stack. This is primarily because at fixed $\mstar$, higher $\dssfr$ implies higher $\halpha$ flux, and thus lower N2. The reason we bin in $\mstar$ and $\dssfr$ is to alleviate the dependence of N2 on $\dssfr$; the difference between our results and those of \citet{Kewley08} effectively reveal the magnitude of this bias. 

We find that the N2 MZR (Figure~\ref{fig:N2_MZR}) asymptotes around solar metallicity and falls slightly below the MZR from \citet{Tremonti04}. This is in agreement with previous studies \citep{Baldwin81,Pettini04} and occurs because nitrogen becomes the dominant coolant at high metallicity, so N2 saturates. At high stellar masses (and metallicities), O3N2 continues to decrease as the intensity of [\ion{O}{iii}] decreases with increasing metallicity. Figures~\ref{fig:dssfr} and~\ref{fig:O3N2_MZR} show that O3N2 begins to flatten at high $\mstar$, but this is likely due to the turnover in the MZR.

In the case of the N2O2 MZR (Figure~\ref{fig:N2O2_MZR}), we note a marginally higher normalization, and significantly larger scatter at fixed $\mstar$, than the other calibrations. This is likely the result of a larger dependence on $\dssfr$. As previously noted, the ionization parameter is likely correlated with $\dssfr$ (see the right panel of Figure~\ref{fig:BPT}). If this is true, the high $\dssfr$ stacks will be biased towards low N2 or high O3N2 \citep{Dopita00,Kewley02,Steidel14}. Given the slope of the strong line calibrations, this will mask the dependence of $\log$(O/H) and $\dssfr$. Being largely insensitive to ionization parameter, N2O2 likely reflects the true relationship between $\log$(O/H) and $\dssfr$.

For most of the $\dssfr$ tracks, the scatter in inferred (O/H) between points is surprisingly small and is much less than that seen in $\ohte$. This is due to the fact that the inferred (O/H) is merely a reflection of how the strong line diagnostics vary as a function of mass. On average, the strong lines exhibit very smooth behavior with mass \citep{Kewley08}. This point was also raised in \citet{Steidel14} and suggests that another parameter other than gas phase oxygen abundance (likely ionization parameter) is tightly coupled to both mass and the strong line ratios. Thus we are able to measure the average strong line value to exquisite precision, but the uncertainty in gas phase oxygen abundance for any one galaxy is set by the scatter in a particular diagnostic ratio at fixed $\mstar$ and $\dssfr$.

\subsection{The $\mzsfrrelation$ Relation}
\label{sec:fmr}

\begin{figure*}
\psfrag{O}[c][][1.]{$\odot$}
\psfrag{S}[c][][1.]{$\star$}
\centering{\includegraphics[scale=1.,width=0.49\textwidth,trim=0.pt 0.pt 270.pt 70.pt,clip]{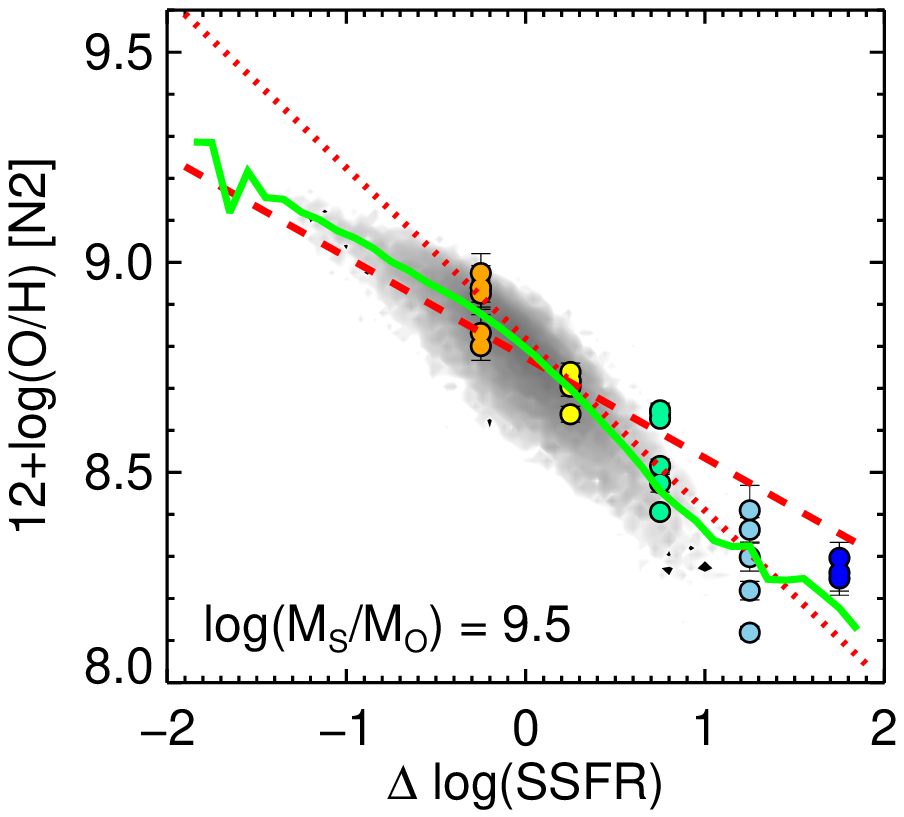}}
\centering{\includegraphics[scale=1.,width=0.49\textwidth,trim=0.pt 0.pt 270.pt 70.pt,clip]{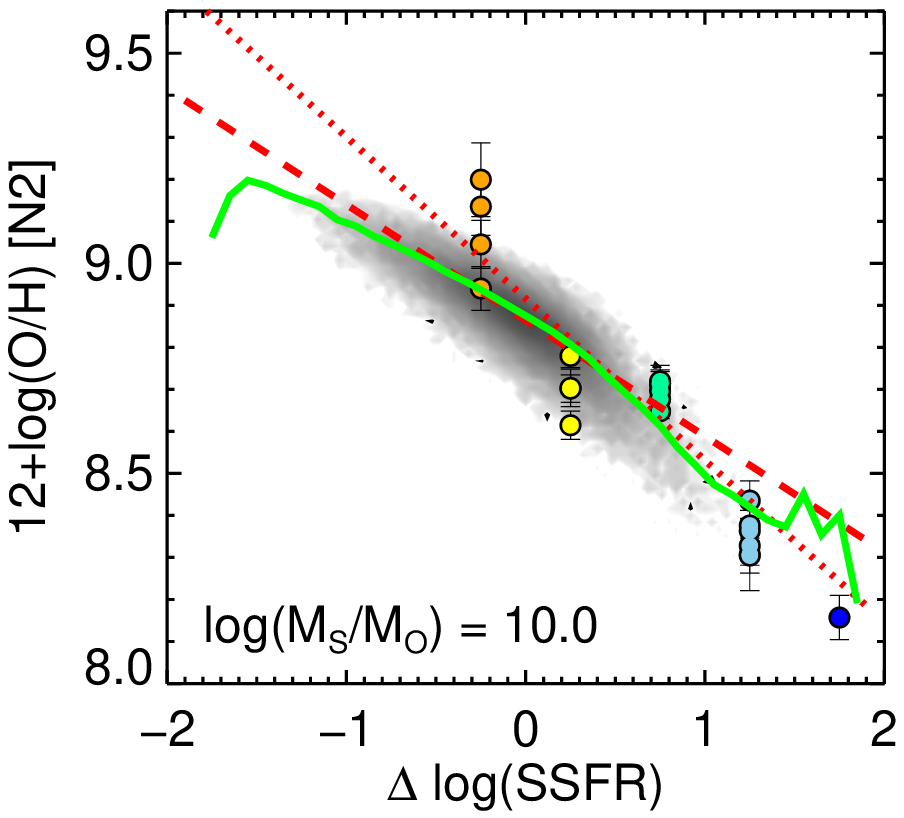}}
\centering{\includegraphics[scale=1.,width=0.49\textwidth,trim=0.pt 0.pt 270.pt 70.pt,clip]{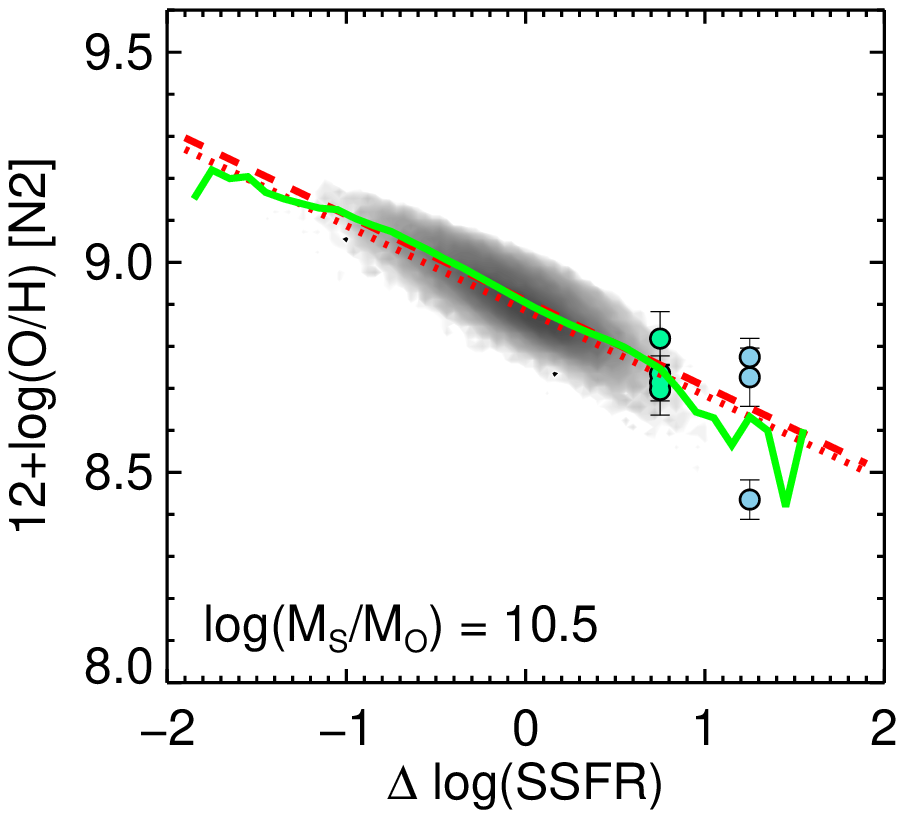}}
\centering{\includegraphics[scale=1.,width=0.49\textwidth,trim=0.pt 0.pt 270.pt 70.pt,clip]{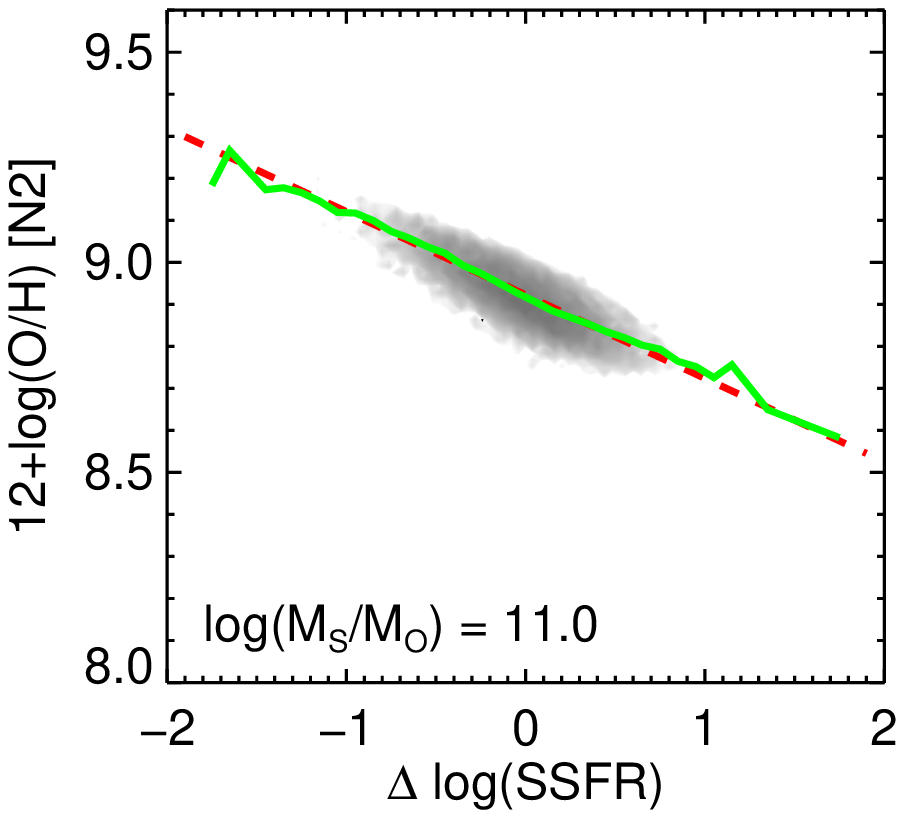}}
\caption{Oxygen abundance as a function of $\dssfr$ for the N2 calibration. Each panel shows galaxies and stacks falling within $\pm 0.25$ dex of the designated mass. The gray contours show the distribution of SDSS galaxies; the circles show the stacks falling within the designated mass range (there are often multiple stacks at a given value of $\dssfr$). The color coding denotes $\dssfr$. Oxygen abundances of the individual galaxies are derived from our new N2 calibration; oxygen abundances of the stacks are derived from the direct method. In each panel, the solid green line shows the median of the individual galaxies, the dashed red line shows the linear fit to the individual galaxies, and the dotted red line shows the linear fit to the stacks. Selection effects cause (1) a reduction in the number of direct method measurements at high $\mstar$, and (2) the fit to the stacks to be biased towards a steeper slopes. This latter effect is due to the fact that the direct method abundances are more easily measured at high $\dssfr$, where the relation between $\log$(O/H) and $\dssfr$ tends to steepen. The nonlinear nature of the green line illustrates the need for such a non-parametric approach. The corresponding plots for the O3N2 diagnostic (not shown) are qualitatively similar to those shown here.} 
\label{fig:salim_comp_n2}
\end{figure*}

\begin{figure*}
\psfrag{O}[c][][1.]{$\odot$}
\psfrag{S}[c][][1.]{$\star$}
\centering{\includegraphics[scale=1.,width=0.49\textwidth,trim=20.pt 0.pt 270.pt 70.pt,clip]{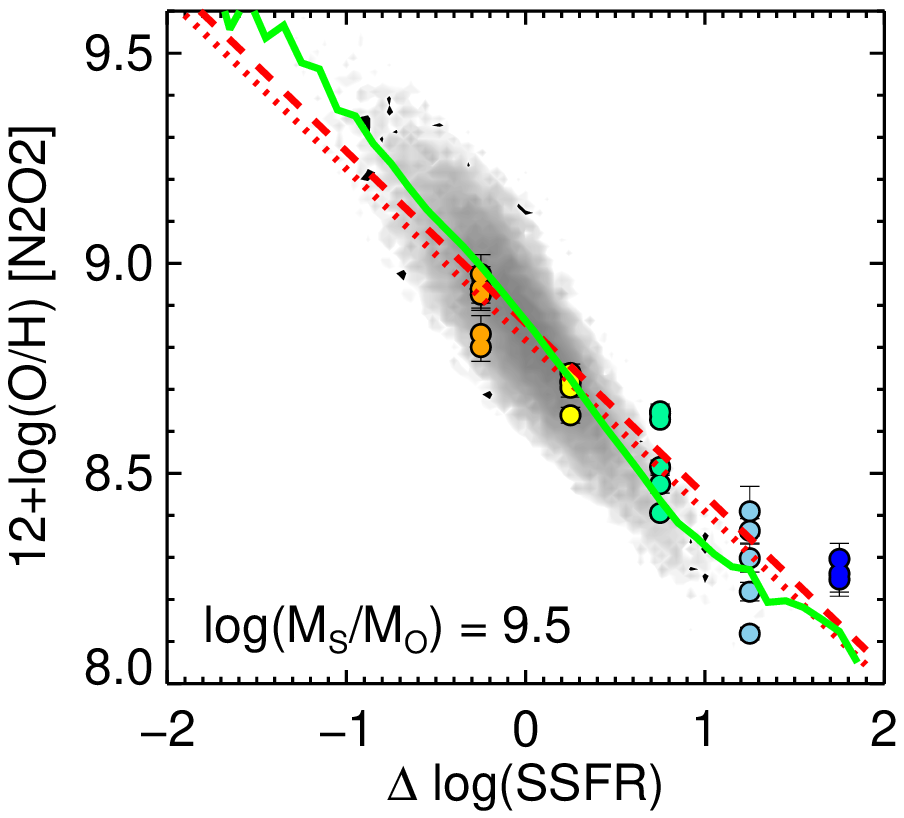}}
\centering{\includegraphics[scale=1.,width=0.49\textwidth,trim=20.pt 0.pt 270.pt 70.pt,clip]{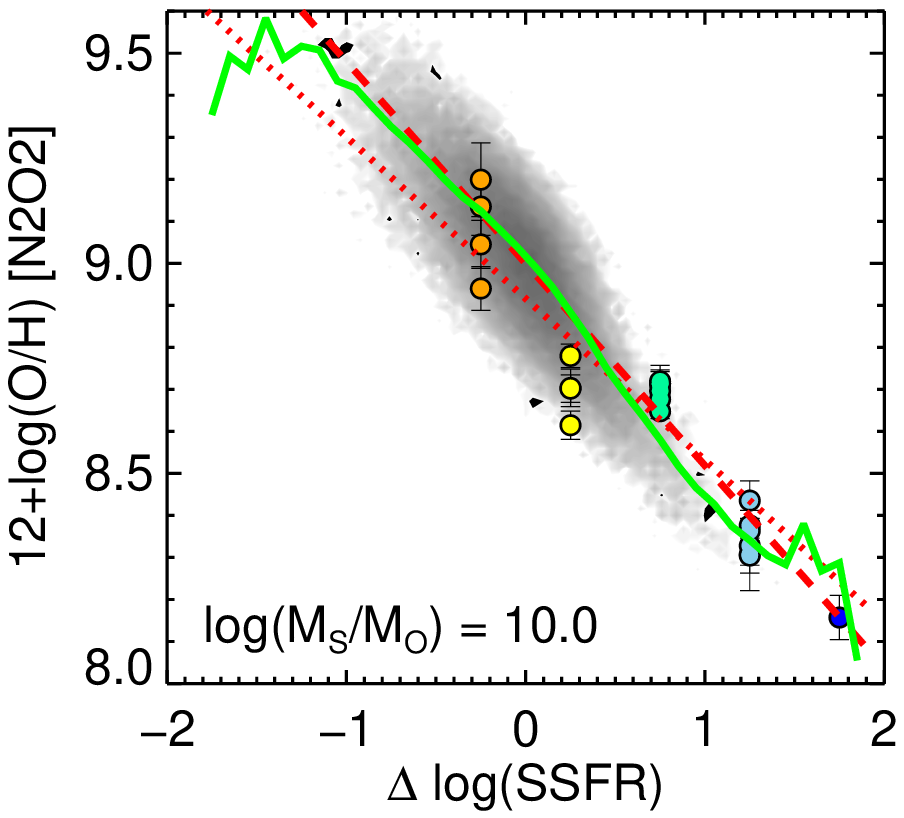}}
\centering{\includegraphics[scale=1.,width=0.49\textwidth,trim=20.pt 0.pt 270.pt 70.pt,clip]{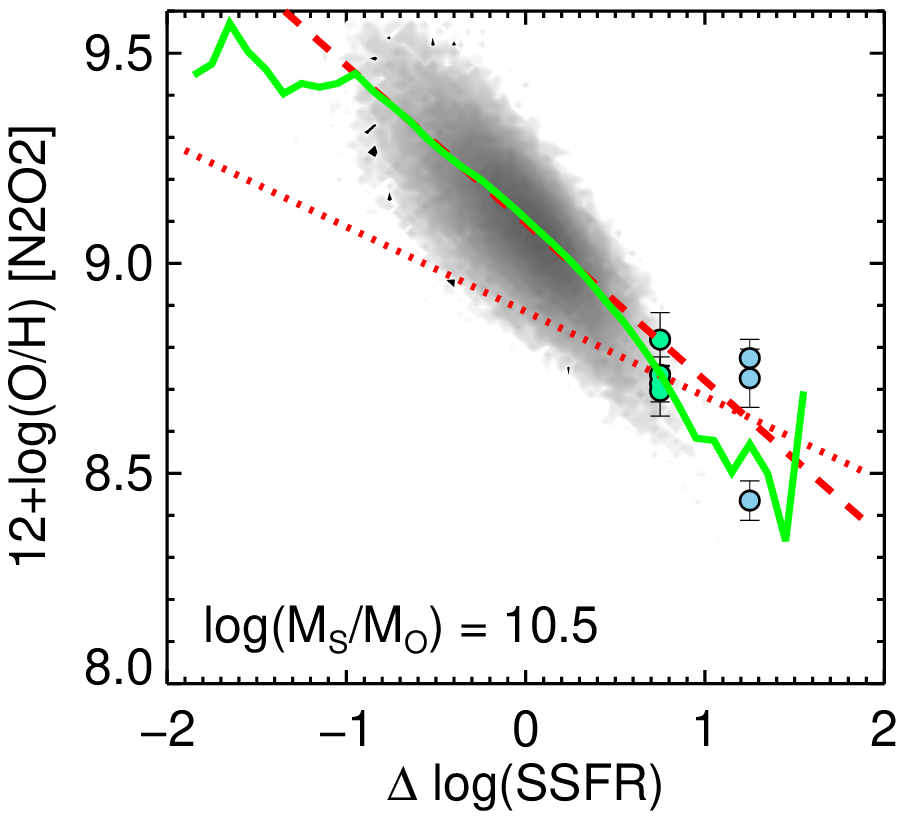}}
\centering{\includegraphics[scale=1.,width=0.49\textwidth,trim=20.pt 0.pt 270.pt 70.pt,clip]{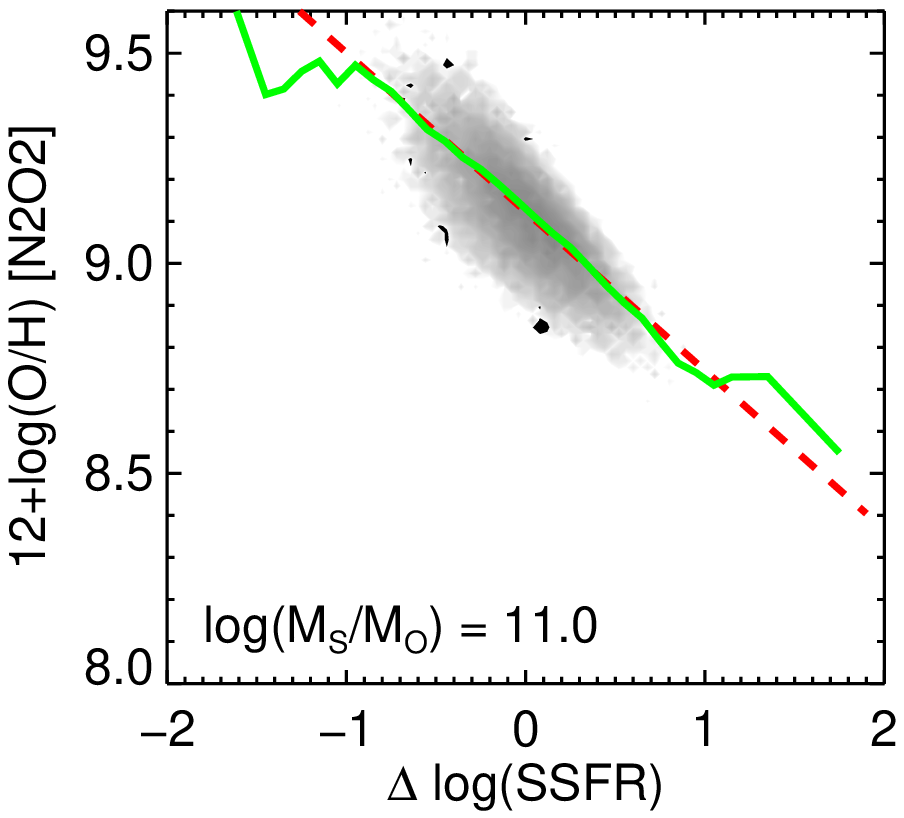}}
\caption{Same as Figure~\ref{fig:salim_comp_n2} but using the N2O2 diagnostic. The dependence on $\dssfr$ is generally steeper than that of the N2 and O3N2 diagnostics, even at high masses. The dependence is also well approximated with the linear parametrization.} 
\label{fig:salim_comp_n2o2}
\end{figure*}

\begin{figure*}
\psfrag{O}[c][][1.]{$\odot$}
\psfrag{S}[c][][1.]{$\star$}
\centering{\includegraphics[scale=1.,width=0.49\textwidth,trim=10.pt 0.pt 270.pt 70.pt,clip]{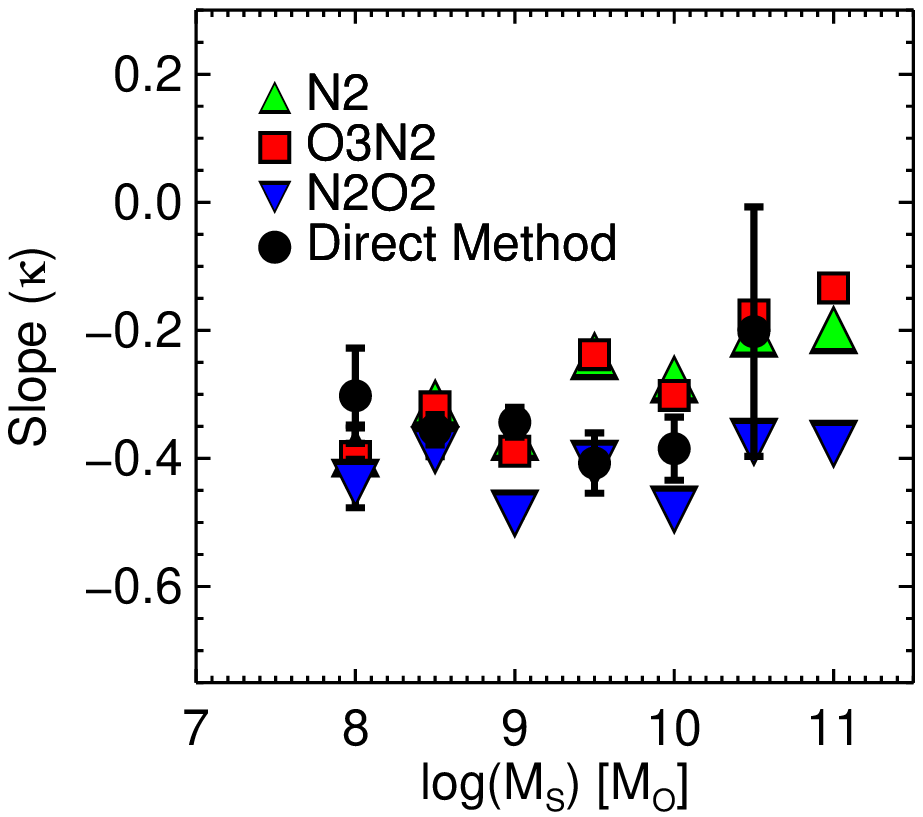}}
\centering{\includegraphics[scale=1.,width=0.49\textwidth,trim=10.pt 0.pt 270.pt 70.pt,clip]{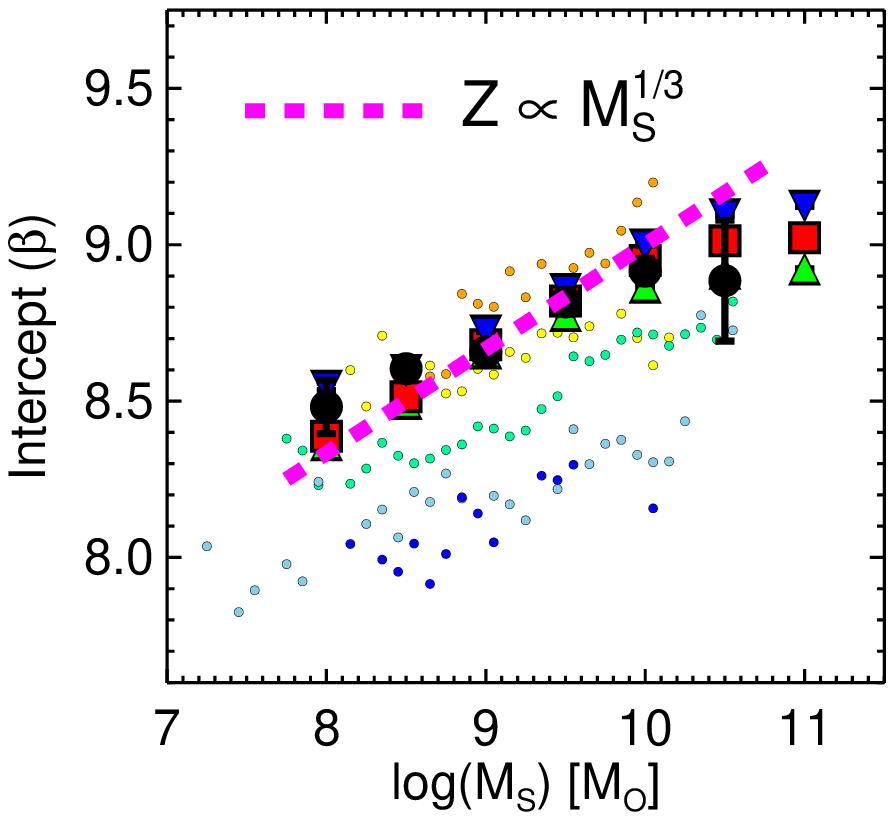}}
\caption{Slopes and intercepts as a function of $\mstar$ for our new strong line calibrations (applied to individual galaxies, green, red, and blue points) and direct method measurements (black points). Left: Slope as a function of $\mstar$. The slopes measured from the strong line calibrations typically agree with those measured from the direct method stacks. The instances of disagreement are likely due to the fact that the direct method abundances are more easily measured at high $\dssfr$, where the relation between $\log$(O/H) and $\dssfr$ appears to steepen. Right: Intercept as a function of $\mstar$. The intercept corresponds to the inferred metallicity of the star forming main sequence. The small circles show where the stacks fall in this parameter space. The stacks are colored according to $\dssfr$. The measured interepts closely track the star forming main sequence, which roughly follows the dashed magenta line corresponding to $Z \propto \mstar^{1/3}$.}
\label{fig:fmrFit}
\end{figure*}

Using the masses and newly derived oxygen abundances of galaxies in the local universe, we can investigate the presence of a Fundamental Metallicity Relation (FMR; \citet{Mannucci10,LaraLopez10}). The formulation of the FMR from \citet{Mannucci10} states that (1) galaxies lie along the projection of the local $\mzsfrrelation$ relation that minimizes the scatter in metallicity, and (2) the relationship is redshift invariant. In this section we will focus on the first of these predictions; we will consider evolution of the $\mzsfrrelation$ relation with redshift in Section~\ref{sec:highz}.

\cite{Salim14} presented a non-parametric analysis framework for investigating the $\mzsfrrelation$ relation in local galaxies. When investigating the nature of the $\mzsfrrelation$ relation, non-parametric techniques are preferred since they do not require a fixed SFR dependence at a given $\mstar$, as is required in the framework of \citet{Mannucci10} or \citet{LaraLopez10}. Following \citet{Salim14, Salim15}, we examine the slope of $12+\log$(O/H) as a function of $\dssfr$ at fixed $\mstar$. For each $\mstar$ bin, we assume the form

\begin{equation}
12+\log({\rm O/H}) = \beta + \kappa*\dssfr. 
\label{eq:slope}
\end{equation}

While this introduces a parametrization, it allows for a direct comparison of the slope $\kappa$ with previous studies \citep[e.g.][]{Salim14,Salim15}. The dependence of $\log$(O/H) on SFR at fixed $\mstar$ is simply $\frac{d\log({\rm O/H})}{d\log({\rm SFR})}\big|_{\small{M_{\star}}} = \frac{d\log({\rm O/H})}{d\Delta \log({\rm SSFR})} = \kappa$. This differs from the parameter $\alpha$ that minimizes the scatter about a surface in $\mzsfrrelation$ space \citep[e.g.,][]{Mannucci10,Yates12,Andrews13}. It is straightforward to convert a value of $\alpha$ to an equivalent value of $\kappa$ if the parametrization of the FMR is known.

\citet{Salim14} find that the slope $\kappa$ is a function of $\mstar$, and becomes flatter at higher masses. They also find that the slope is a function of $\dssfr$, and becomes steeper at higher $\dssfr$.  We apply their framework to the galaxies in our sample. We measure $\dssfr$ with Equation~\ref{eq:dssfr}, and apply our new strong line calibrations to derive oxygen abundances.

Each panel of Figure~\ref{fig:salim_comp_n2} shows $\log$(O/H) as a function of $\dssfr$ for a given $\mstar$ denoted in the bottom left corner. We include all galaxies and stacks with masses that fall within the $\pm 0.25$ dex $\mstar$ window of each panel. The circles show the direct method abundances of the stacks. The stacks are 0.10 dex wide in $\mstar$, so there are multiple stacks at fixed $\dssfr$ within the $\mstar$ window of each panel. The gray contours show the SDSS galaxies with oxygen abundances determined with our new calibration.

For each $\mstar$, we fit $\log$(O/H) as a function of $\dssfr$ per Equation~\ref{eq:slope}. The dashed red lines show the fits resulting from the SDSS galaxies; the dotted red lines show the fits to the stacks. Note that for higher masses ($\log(\mstar/M_{\odot}) \gtrsim 10.0$) there are few to no stacks with direct method abundances. While in some cases the slopes derived from the direct method differ from those derived from the individual galaxies (e.g., $\log(\mstar/M_{\odot}) = 9.5$), we typically find agreement within the error bars. 

The solid green line in each panel shows the median $\log$(O/H) as a function of $\dssfr$. The relationship between $\log$(O/H) and $\dssfr$ is non-linear and appears to steepen at high $\dssfr$, particularly for the lower mass bins. This is in agreement with \citet{Salim14} and illustrates the need for a non-parametric approach when investigating the $\mzsfrrelation$ relation. Since our detection of auroral lines is biased towards high $\dssfr$, we have relatively more direct method measurements at high $\dssfr$, which effectively biases the fit to the stacks towards a steeper slope. Accounting for $\dssfr$ does lead to a reduction in scatter; the scatter in (O/H) at fixed $\mstar$ and $\dssfr$ is somewhat lower than the scatter at fixed $\mstar$ alone. In the case of N2, the scatter in O/H at fixed $\mstar$ is $\sim 0.12$, while the scatter around the running median is $0.07$.

We perform this analysis for the O3N2 and N2O2 diagnostics as well. The results for the O3N2 diagnostic are qualitatively similar to those of the N2 diagnostic. In Figure~\ref{fig:salim_comp_n2o2} we examine the results of this non-parametric approach with the N2O2 diagnostic. The green line shows the median $\log$(O/H) of the individual galaxies, while the dashed (dotted) red lines show the parametrized fit to the slope of the galaxies (stacks). Interestingly, the N2O2 diagnostic removes much of the nonlinearity of the relationship between $\log$(O/H) and $\dssfr$; the green and red lines agree across a wide range of $\dssfr$. Furthermore, the slope remains relatively steep, even at high masses, which is not the case for the other diagnostics.

The results of the linear fits for each diagnostic are shown in Figure~\ref{fig:fmrFit}. The left panel shows the measured slope (for both the SDSS galaxies and direct method stack abundances). The right panel shows the corresponding intercept for each fit; the small circles show where the stacks fall in the $\mstar-Z$ plane. The measured intercepts (right panel) closely track the star forming main sequence, which also follows the $Z \propto \mstar^{1/3}$ scaling denoted by the dashed magenta line. This is consistent with momentum driven winds and a mass loading parameter $\eta$ which scales approximately as $\eta \propto \mstar^{-1/3}$ \citep{Murray05,Oppenheimer06}. 

The left panel of Figure~\ref{fig:fmrFit} presents clear evidence for evolution of the slope $\kappa$ as a function of $\mstar$ for the N2 and O3N2 diagnostics. The slope is steeper at lower masses, in agreement with previous studies \citep{Ellison08Apj,Salim14}. We measure $\kappa\sim -0.2$ to $-0.4$. \citet{Andrews13} measured $\alpha = 0.66$ and the slope of the FMR to be 0.43 with the direct method. Converting their direct method $\alpha$ to an equivalent value of $\kappa$ yields $\kappa \sim -0.28$, which is in good agreement with our measurements. Furthermore, the tension between the slope derived from direct method abundances and that derived from strong line inferred abundances is significantly reduced from that found in \citet{Andrews13}. Our values of $\kappa$ are on average steeper than \citet{Salim14} found. This is at least in part due to the fact that our new calibrations incorporate $\dssfr$ explicitly.

The nonlinear dependence of $\log$(O/H) on $\dssfr$ is most prominent in the low mass panels of Figure~\ref{fig:salim_comp_n2}. There is a break in slope between $\log$(O/H)$_{\rm N2}$ and $\dssfr$, which appears to denote a boundary between highly star forming galaxies and more moderately star forming galaxies. \citet{Salim14, Salim15} interperet this break and the general flattening of the slope with $\mstar$ in the context of models from \citet{Zahid14}. They suggest that the ISM of the more evolved galaxies is saturated and thus the gas phase abundances are largely insensitive to inflows of pristine gas and the resulting variations in $\dssfr$. In contrast, the more vigorously star forming galaxies have lower gas phase abundances which are more sensitive to inflows of pristine gas. However, the flattening in slope could also be due to the N2 diagnostic losing sensitivity at high metallicities. This would not, however, explain the similar behavior seen for the O3N2 diagnostic (see Figure~\ref{fig:fmrFit}) which is expected to remain sensitive to oxygen abundance in the high metallicity regime.

The break in slope is not present in Figure~\ref{fig:salim_comp_n2o2} for N2O2. Furthermore, the slope in Figure~\ref{fig:salim_comp_n2o2} is relatively steep and constant for all $\mstar$. Since the N2O2 diagnostic is insensitive to ionization parameter, this may mean that the ionization parameter is more tightly coupled to $\dssfr$ in intensely star forming galaxies. For instance, suppose an increase in SFR in a highly star forming galaxy produced a larger increase in ionization parameter than in a more moderately star forming galaxy with the same stellar mass. This would bias the N2 and O3N2 diagnostics in the direction of lower metallicity and cause the slope between inferred $\log$(O/H) and $\dssfr$ to steepen. This would explain why the break is present for N2 and O3N2, but not N2O2. We emphasize that Figures~\ref{fig:salim_comp_n2} and~\ref{fig:salim_comp_n2o2} show how changes in $\mstar$ and $\dssfr$ affect the diagnostics, from which we only \textit{infer} a metallicity. While the break in slope may be a real effect resulting from the physical processes governing the $\mzsfrrelation$ relation, there remain potential biases associated with strong line calibrations. 

\subsection{Application of New Calibrations to High Redshift Galaxies}
\label{sec:highz}

\begin{figure*}
\psfrag{O}[c][][1.]{$\odot$}
\psfrag{S}[c][][1.]{$\star$}
\centering{\includegraphics[scale=1.,width=0.49\textwidth,trim=10.pt 0.pt 260.pt 70.pt,clip]{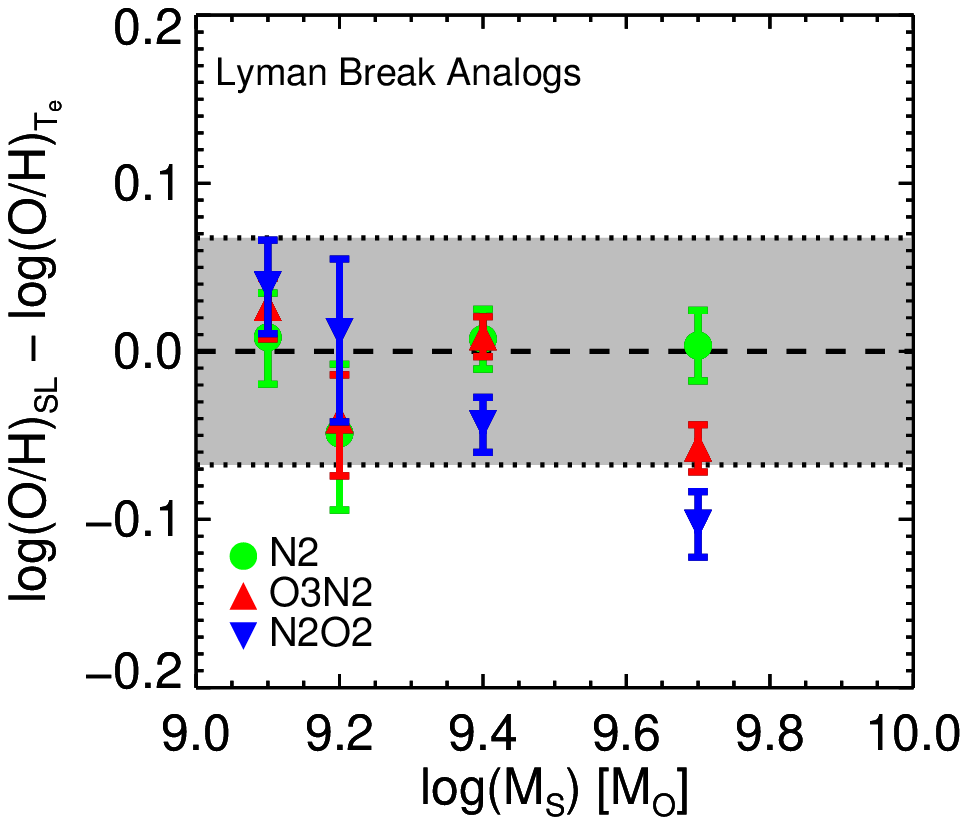}}
\centering{\includegraphics[scale=1.,width=0.49\textwidth,trim=10.pt 0.pt 260.pt 70.pt,clip]{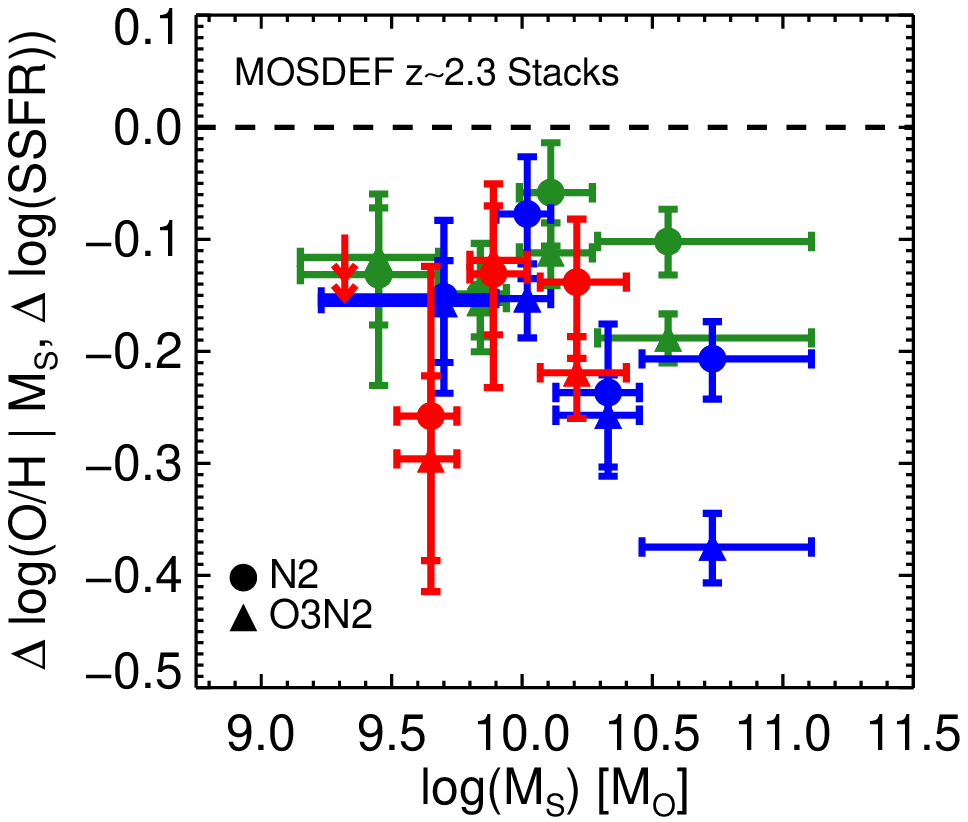}}
\caption{Left: Deviation in strong line (SL) inferred (O/H) from the direct method (O/H) for the 4 LBAs ($z \sim 0.2$) from \citet{Brown14}. The error bars are generated from the Monte Carlo technique previously described (see Section~\ref{sec:abund}). The gray band shows the average uncertainty in the direct method oxygen abundances. The oxygen abundances determined with our new strong line calibrations are generally consistent with the measured direct method abundances. Right: Deviation in (O/H) of high redshift galaxies from local galaxies of the same mass and SFR. The MOSDEF z$\sim$2.3 points are from the \citet{Sanders15} stacks. The blue, red, and green denote their high, low, and composite SFR stacks respectively. Both our N2 (circles) and O3N2 (triangles) calibrations are shown. The error bars in the mass direction show the range of $\mstar$ for each bin and the error bars in $\Delta$ (O/H) are generated from the Monte Carlo technique previously described (see Section~\ref{sec:abund}). The high redshift galaxies have lower (O/H) than local galaxies of the same $\mstar$ and SFR regardless of mass, SFR, or strong line diagnostic.}
\label{fig:SL_highz}
\end{figure*}

Most galaxies found in high redshift surveys are qualitatively similar to gas rich, metal poor, highly star forming galaxies in the local universe \citep{Steidel14,Kriek14,Shapley15,delosReyes15}. This is at least in part a selection effect. At high redshift, bright emission line galaxies are easier to detect than quiescent galaxies. However, the average SFR and SSFR of the universe does indeed increase with redshift, peaking near $z \sim 2$ \citep[e.g., see the compilation by][]{AHopkins06}. In this section we investigate how the mean properties of high redshift galaxies compare to those of local star forming galaxies, as well as whether or not the diagnostic tools developed from galaxies in the local universe can yield useful information when applied to high redshift galaxies.

\subsubsection{Are the Calibrations Valid at High Redshift?}

The calibrations derived in Section~\ref{sec:results} incorporate $\mstar$ and $\dssfr$ relative to the local star forming main sequence. When applying these calibrations to high redshift galaxies there is an implicit comparison to the local star forming main sequence, rather than the star forming main sequence of the high redshift universe. Since the average star formation rate of the universe evolves with redshift, so does the star forming main sequence. In this sense, the local star forming main sequence is a somewhat arbitrary (albeit convenient) zero point for our calibrations. Utilizing a $\dssfr$ defined relative to the high redshift star forming main sequence would require recalibrating the diagnostics using high redshift galaxies. This would merely amount to a zero-point shift \citep{Salim15}, since in our framework the higher (S)SFRs would be balanced by lower metallicities.

One possible concern is whether or not it is appropriate to apply our calibrations to high redshift galaxies. \citet{Steidel14} argue that the position of high redshift galaxies in the BPT diagram is largely independent of (O/H), and primarily determined by the ionization parameter $\Gamma$, which is highly dependent on $\teff$, the density of star formation, and geometrical effects. They find that the correlation between (O/H) and the strong line ratios is most likely a result of the correlation between $\Gamma$, $\teff$, and the stellar metallicity which, for young stellar populations, reflects the gas phase metallicity. The average $\teff$ may indeed evolve with redshift due to the compact, gas rich, low metallicity environments that become more common at higher redshifts. These conditions could result in stellar populations with abnormally hard ionizing spectra that drive unusual ionization conditions and abundances \citep{Eldridge09,Brott11,Levesque12,Kudritzki00,Kewley13a}. \citet{Steidel14} show that a factor of 2.5 change in $\Gamma$ has the same order of magnitude effect on N2 as a factor of five change in $Z$. Even in the local universe, a factor of 2.5 variation in ionization parameter from one object to another is not unreasonable \citep{Zahid12a}, although the $z\sim 2.3$ galaxies would require a \textit{systematic} increase in ionization parameter of this order of magnitude. While there is evidence that the ionization conditions of high redshift galaxies are similar to local \ion{H}{ii} regions \citep{Nakajima13}, the validity of local strong line calibrations at high redshift is further complicated by the fact that the abundance of nitrogen relative to oxygen may increase with redshift \citep{Steidel14,Masters14}.

While we do not yet have direct method oxygen abundances for a large sample of $z \geq 2$ galaxies, \citet{Brown14} measured the direct method oxygen abundances and strong line ratios of several Lyman Break Analogs \citep[LBAs;][]{Heckman05, Hoopes07, BasuZych07, Overzier08, Overzier09, Overzier10, Goncalves10}. LBAs are local ($z \sim 0.2$) versions of the Lyman Break Galaxies which dominated the SFR of the universe at $z \gtrsim 2.5$ \citep[for a review of LBGs, see][]{Giavalisco02}. In the left panel of Figure~\ref{fig:SL_highz} we compare the oxygen abundance determined with our new calibrations with the direct method (O/H) for the four LBAs from \citet{Brown14}. The circles, triangles, and inverted triangles denote the deviation of the inferred (O/H) from the direct method (O/H) for our N2, O3N2, and N2O2 calibrations respectively. The gray shaded region shows the average uncertainty of the direct method measurements. 

The choice of star formation rate indicator is a source of systematic error. Our calibrations are derived using the SFRs from the $\mpa$ pipeline. In order to minimize systematic effects associated with the SFR of LBAs, we adopt the SFRs from the $\mpa$ catalog, which agree with the $\halpha$ derived SFRs from \citet{Overzier09}. While the $\halpha+24\mu$m SFRs from \citet{Overzier09} are regarded as the optimal SFR indicator, these values are systematically high compared to the $\halpha$ derived SFRs and result in correspondingly low oxygen abundances. Thus we recommend $\halpha$ derived SFRs when applying these calibrations.

In general, the oxygen abundances predicted by our new calibrations and the direct method oxygen abundances for these LBAs agree quite well. The biggest difference is the N2O2 based metallicity of the most massive LBA from \citet{Brown14}, J005527, which is 1$\sigma$ larger than the direct method metallicity. However, this object displays features consistent with Wolf-Rayet stars, which may drive unusual (N/O) ratios \citep{Pagel86,Henry00,Brinchmann08,LopezSanchez10,Berg11}. We conclude that our new calibrations are suitable for use in LBAs, and that our new calibrations will produce reliable oxygen abundance estimates in the high redshift universe if the ionization conditions of LBAs are representative of their high-$z$ counterparts. Nevertheless, direct method abundance measurements for high redshift galaxies are still needed to determine if local calibrations are suitable for high redshift galaxies.

\subsubsection{Application to MOSDEF $z\sim2.3$ Galaxies}

The MOSFIRE Deep Evolution Field (MOSDEF) survey \citep{Kriek14} is a spectrocopic survey investigating the rest frame optical emission lines of high redshift star forming galaxies. \citet{Sanders15} used a sample of MOSDEF galaxies to stack spectra in $\mstar$ and $\mstar$--SFR bins in order to measure the rest frame optical emission lines of $z\sim2.3$ galaxies with high precision. We use the published $\mstar$, SFR, and emission line data from \citet{Sanders15} to calculate $\dssfr$ relative to the local star forming main sequence. We apply our new strong line calibrations to the high and low SFR stacks from \citet{Sanders15} (shown as crosses in Figures~\ref{fig:N2_MZR} and~\ref{fig:O3N2_MZR}). We determine the uncertainty in oxygen abundances using a Monte Carlo technique similar to that used to determine the uncertainties in our own abundances (see Section~\ref{sec:abund}). The error bars in the $\mstar$ direction show the mass range of galaxies in the stack. These galaxies fall well below the local MZR. This is in agreement with \citet{Sanders15}, and other studies which have shown that high redshift, highly star forming galaxies tend to have low gas phase oxygen abundances \citep[e.g.][]{Erb06,Maiolino08,Maier14}.

Conceptually, if the gas fueling the star formation has low metallicity, then the ISM of highly star forming galaxies will be relatively metal poor \citep{Ellison08Apj, Mannucci10, LaraLopez10}. However, Figures~\ref{fig:N2_MZR} and~\ref{fig:O3N2_MZR} also show that high redshift galaxies from \citet{Sanders15} are metal poor relative to our low-$z$ stacks with similar $\mstar$ and SFR. The right panel of Figure~\ref{fig:SL_highz} shows a quantitative comparison of where high redshift galaxies fall relative to local galaxies with similar $\mstar$ and SFR. We find that the high redshift galaxies from \citet{Sanders15} have metallicities that are on average $\sim0.1-0.2$ dex lower than local galaxies of the same $\mstar$ and SFR. There is also evidence that the offset in $\log$(O/H) increases with $\mstar$, as noted in \citet{Salim15}. This trend holds for both N2 and O3N2. We did not apply our N2O2 calibration as the [\ion{O}{ii}]~$\lambda$3727 \AA\ line does not fall within the spectral range of the MOSFIRE data reported by \citet{Sanders15}. The offset of the \citet{Sanders15} galaxies toward lower oxygen abundances than local galaxies with the same $\mstar$ and SFR appears to contradict the existence of an FMR, and requires some redshift dependence of the $\mzsfrrelation$ relation.

\citet{Zahid14} use analytic and numerical models to quantify the evolution in their datasets. Their model, which they refer to as the Universal Metallicity Relation (UZR), assumes all galaxies evolve along the star forming main sequence. They model the MZR at any epoch as 

\begin{equation}
12+\log({\rm O/H}) = Z_O + \log\left[ 1 - \exp\left(-\left[\frac{\mstar}{M_O}\right]^{\gamma}\right)\right].
\label{eq:uzr}
\end{equation}

\noindent They find that the shape of the MZR is constant (i.e. universal). Only the characteristic turnover mass $\mturn$ increases with redshift such that at fixed $\mstar$, O/H decreases with redshift. Above $\mturn$, galaxies have essentially the same metallicity $\zmax$. 

\citet{Salim15} suggest that the high metallicities act as a buffer against inflows diluting the ISM, resulting in the break in $\kappa$ seen in the top panels of Figure~\ref{fig:salim_comp_n2}. With a sufficiently large sample of high redshift galaxies resolving the turnover in the MZR, it may be possible to directly test the evolution of $\mturn$ with redshift within the framework of Section~\ref{sec:fmr}. If $\mturn$ increases with redshift as argued by \citet{Zahid14}, the break in $\kappa$ should occur at a higher mass than observed for local samples of galaxies. \citet{Salim15} examine the $\mzsfrrelation$ relation with the high redshift galaxies from \citet{Steidel14}, as well as local galaxies with relatively high values of $\dssfr$. Their results suggest that $\kappa$ flattens at \textit{high} $\dssfr$, but current samples of high redshift galaxies are not yet complete enough to reveal a break in $\kappa$ at lower values of $\dssfr$.


\section{Summary}
\label{sec:conc}

We have recalibrated strong line diagnostics with direct method oxygen abundances of galaxies and applied the new calibrations to investigate the $\mzsfrrelation$ relation. We stacked $\sim 2 \times 10^5$ spectra of star forming galaxies in the local universe in $\mstar$ and offset from the star forming main sequence. Our main results are:

\begin{itemize}
  \item We recalibrated the relationship between $\ohte$ and the N2, O3N2, N2O2 strong line ratios. This included incorporation of $\dssfr$ as an additional parameter.
  \item For the N2 and O3N2 diagnostics we find a higher (O/H) normalization, but similar slope, as previous calibrations. We attribute this difference to the fact that previous calibrations are based on individual \ion{H}{ii} regions. No single calibration significantly outperforms the others. The O3N2 diagnostic is the most accurate of the three for 43\% (47/110) of the stacks, but N2O2 is typically a close second and subject to fewer biases. 
  \item We apply our new calibrations to local star forming galaxies. In the context of galaxy evolution models, our result that the slope of our new calibrations is similar to previous calibrations implies the scaling of galactic outflows with stellar mass remains unchanged.
  \item We adopt the non-parametric framework presented in \citet{Salim14} to investigate the $\mzsfrrelation$ relation in the local universe. When using the N2 and O3N2 diagnostics we find variation in the SFR dependence with both $\mstar$ and $\dssfr$, as noted in previous studies. The N2O2 diagnostic produces a nearly constant slope, independent of $\mstar$ and $\dssfr$. Below $\log(\mstar/\msol) \sim 10$, the slopes measured with strong line diagnostics are in agreement with each other and consistent with the direct method slope to within $\sim10\%$. At higher masses, the uncertainty in the direct method slope increases significantly, and the N2 and O3N2 inferred slopes flatten compared to N2O2. We note a modest reduction of scatter in $\log$(O/H) at fixed $\mstar$ and $\dssfr$.
  \item We also apply our new calibrations to high redshift galaxies presented in \citet{Sanders15}. We find these galaxies to be systematically metal poor compared to local galaxies of the same $\mstar$ and SFR, and conclude the $\mzsfrrelation$ relation evolves with redshift.
  \item It is possible that our O/H estimates of high redshift galaxies are biased by the ionization conditions of the high redshift universe. While direct method measurements of high redshift galaxies are required to definitively test if this is the case, we apply our new calibrations to the LBAs from \citet{Brown14} and find consistent results with the direct method measurements of those systems.
\end{itemize}

There remains some degree of uncertainty as to whether or not these calibrations are valid in the high redshift universe. The ideal path forward would be to recalibrate these empirical relations at $z \sim 2.3$. While direct method oxygen abundance determinations at high redshift are challenging, recent progress has been made. There have been several direct method abundance measurements obtained at $z \sim 1$ \citep{Hoyos05, Kakazu07, Amorin10, Amorin12}, and \citet{Yuan09} used gravitational lensing to measure $\oaur$ at $z\sim1.7$. Most recently, \citet{Jones15} showed that $\alpha$ element strong line abundance diagnostics are reliable up to at least $z \sim 0.8$. Additionally, \citet{Steidel14} report that direct method oxygen abundances (in addition the [\ion{O}{ii}], [\ion{O}{iii}], $\halpha$, $\hbeta$, [\ion{N}{ii}], and [\ion{S}{ii}] optical strong lines) will soon be available for a subset of the KBSS-MOSFIRE targets at $z \approx 2.36-2.57$. This will improve constraints on the $\mzsfrrelation$ relation and ionization conditions in the early universe.    

While we have restricted ourselves to two applications of our newly derived calibrations (the $\mzsfrrelation$ relation and the high redshift universe), there are many other potential applications of these calibrations. For example, a set of abundance diagnostics based on direct method abundances of galaxies rather than individual \ion{H}{ii} regions is invaluable for any study concerned with gas phase abundances of galaxies, such as transient surveys like ASASSN \citep{Shappee14} and ZTF \citep{Bellm14}. There are also many applications to IFU spectroscopic galaxy surveys \citep[e.g. MaNGA,][]{Bundy15}, particularly in regions of galaxies where the weak lines are not detected. Lastly, next generation galaxy surveys like DESI \citep{DESI} will be able to make use of these calibrations to study much larger samples of galaxies. 

\section*{Acknowledgements}
We thank Roberto Maiolino, Gwen Rudie, Samir Salim, Ryan Sanders, and Chuck Steidel for comments on an early draft. We also appreciate many helpful comments and suggestions by the referee.

We appreciate the MPA-JHU group for making their catalog publicly available.

The STARLIGHT project is supported by the Brazilian agencies CNPq, CAPES and FAPESP and by the France-Brazil CAPES/Cofecub program.

Funding for the SDSS and SDSS-II has been provided by the Alfred P. Sloan Foundation, the Participating Institutions, the National Science Foundation, the U.S. Department of Energy, the National Aeronautics and Space Administration, the Japanese Monbukagakusho, the Max Planck Society, and the Higher Education Funding Council for England. The SDSS Web Site is http://www.sdss.org/.

The SDSS is managed by the Astrophysical Research Consortium for the Participating Institutions. The Participating Institutions are the American Museum of Natural History, Astrophysical Institute Potsdam, University of Basel, University of Cambridge, Case Western Reserve University, University of Chicago, Drexel University, Fermilab, the Institute for Advanced Study, the Japan Participation Group, Johns Hopkins University, the Joint Institute for Nuclear Astrophysics, the Kavli Institute for Particle Astrophysics and Cosmology, the Korean Scientist Group, the Chinese Academy of Sciences (LAMOST), Los Alamos National Laboratory, the Max-Planck-Institute for Astronomy (MPIA), the Max-Planck-Institute for Astrophysics (MPA), New Mexico State University, Ohio State University, University of Pittsburgh, University of Portsmouth, Princeton University, the United States Naval Observatory, and the University of Washington.

\bibliography{calib}
\bsp	
\label{lastpage}
\end{document}